%% file: nnbar.tex
\documentclass[a4paper,11pt]{article}
\pdfoutput=1 

\usepackage{jinstpub} 

\usepackage{lineno}

\usepackage{siunitx}
\usepackage{array}

\usepackage{caption}
\usepackage{makecell}
\usepackage{algorithm}
\usepackage{pm-isomath}
\usepackage{comment}
\usepackage{float}
\usepackage{siunitx}
\usepackage[version=4]{mhchem}
\DeclareSIUnit\angstrom{\text {Å}}
\usepackage{booktabs}
\usepackage{multicol}
\usepackage{multirow}
\usepackage{cleveref}
\usepackage[flushleft]{threeparttable}
\usepackage{tikz} 
\usetikzlibrary{shapes.geometric, arrows} 
\usepackage[section]{placeins} 
\newcommand{\software}[1]{{\fontfamily{qcr}\selectfont {#1}}} 
\usepackage{caption} 
\usepackage{subcaption} 

\title{The Development of the NNBAR Experiment}

\author[a]{F.~Backman,}
\author[b,c]{J.~Barrow,}
\author[d]{Y.~ Be{\ss}ler,}
\author[e]{A.~Bianchi,}            %
\author[a]{C.~Bohm,}
\author[f]{G.~Brooijmans,}
\author[g]{H.~Calen,}
\author[h]{J.~Cederk\"{a}ll,}
\author[e]{J.~I.~M.~Damian,}
\author[i,j]{E.~Dian,}    
\author[e]{D. D. Di Julio,}
\author[a]{K.~Dunne,}
\author[g]{L.~Eklund,}
\author[e]{M.~J.~Ferreira,}      %
\author[k]{P.~Fierlinger,}
\author[e]{U.~Friman-Gayer,}      %
\author[d]{C.~Happe,}
\author[e]{M.~Holl,}
\author[g]{T.~Johansson,} 
\author[l]{Y.~Kamyshkov,} 
\author[m]{E.~Klinkby,}            %
\author[n]{R.~Kolevatov,}  %
\author[g]{A.~Kupsc,} 
\author[a,h]{B.~Meirose,}
\author[a]{D.~Milstead,} 
\author[o]{A.~Nepomuceno,}
\author[p]{T.~Nilsson,}
\author[h]{A.~Oskarsson}     %
\author[h]{H.~Perrey}        %
\author[e]{K.~Ramic,}       %
\author[e]{B.~Rataj,} 
\author[m]{N.~Rizzi,} 
\author[e,h]{V.~Santoro,}  
\author[a]{S.~Silverstein,} 
\author[q,r,s]{W.M.~Snow,} 
\author[e]{A.~Takibayev }          %
\author[t]{R.~Wagner}              %
\author[g]{M.~Wolke,} 
\author[a]{S.C.~Yiu,}  
\author[u]{A.R.~Young,} 
\author[e]{L.~Zanini,} 
\author[t]{O.~Zimmer,}

\affiliation[a]{Department of Physics, Stockholm University, 106 91 Stockholm, Sweden}
\affiliation[b]{Massachusetts Institute of Technology, Dept. of Physics, Cambridge, MA 02139, USA}
\affiliation[c]{School of Physics and Astronomy, Tel Aviv University, Tel Aviv 69978, Israel}
\affiliation[d]{Forschungszentrum J\"ulich, 52425 J\"ulich
Germany}
\affiliation[e]{European Spallation Source ERIC, Partikelgatan 5, 22484 Lund, Sweden}
\affiliation[f]{Department of Physics, Columbia University, New York, NY 10027, United States of America}
\affiliation[g]{Department of Physics and Astronomy, Uppsala University, Uppsala, Sweden}
\affiliation[h]{Department of Physics, Lund University, P.O. Box 118, SE-221 00 Lund, Sweden}
\affiliation[i]{Mirrotron Ltd., 29-33 Konkoly Thege Mikl\'{o}s \'{u}t, 1121 Budapest, Hungary}
\affiliation[j]{Centre for Energy Research, 29-33 Konkoly Thege Mikl\'{o}s \'{u}t, 1121 Budapest, Hungary}
\affiliation[k]{Physikdepartment Technische Universität München James-Franck-Str. 1 85748 Garching Deutschland}
\affiliation[l]{Department of Physics and Astronomy, The University of Tennessee, Knoxville, TN 37996, USA}
\affiliation[m]{DTU Physics, Technical University of Denmark, Frederiksborgvej 399, DK-4000 Roskilde, Denmark}
\affiliation[n] {European Spallation Source Consultant, Norway}
\affiliation[o]{Departamento de Ci\^encias da Natureza, Universidade Federal Fluminense, Rua Recife, 28890-000 Rio das Ostras, RJ, Brazil}
\affiliation[p]{Institutionen f{\"o}r Fysik, Chalmers Tekniska H\"{o}gskola, Sweden}
\affiliation[q]{Department of Physics, Indiana University, 727 E. Third St., Bloomington, IN, USA, 47405}
\affiliation[r]{Indiana University Center for Exploration of Energy \& Matter, Bloomington, IN 47408, USA}
\affiliation[s]{Indiana University Quantum Science and Engineering Center, Bloomington, IN 47408, USA}
\affiliation[t]{Institut Laue-Langevin, 71 Avenue des Martyrs, 38042 Grenoble, France}
\affiliation[u]{Department of Physics, North Carolina State University, Raleigh, NC 27695-8202, USA}

\emailAdd{valentina.santoro@ess.eu}

\abstract{ The NNBAR experiment for the European Spallation Source will search for free neutrons converting to antineutrons with a sensitivity improvement of three 
orders of magnitude compared to the last such search. This paper describes progress towards a conceptual design report for NNBAR. The design of a moderator, neutron reflector, beamline, shielding and annihilation detector is reported. The simulations used form part of a model which will be used for optimisation of the experiment design and quantification of its sensitivity.     

}


\begin{document}
\maketitle
\flushbottom

\input{introduction}

\input{ess}
\input{moderator}

\input{reflector_shortened}
\input{magnetics}

\input{detector}
\input{summary}
\input{acknowledgements}








\input{nnbar.bbl}


\end{document}

%% file: introduction.tex
\section{Introduction}

The European Spallation Source (ESS), currently under construction in Lund, Sweden~\cite{Peggs:2013sgv}, will be the world's brightest neutron source.  The facility's unique capabilities will exceed and complement those of today's leading neutron sources. Taking advantage of this unique potential of the ESS, the HIBEAM/NNBAR collaboration has proposed a two-stage program of experiments~\cite{Addazi:2020nlz} to perform high precision searches for neutron conversions in a range of baryon number violation (BNV) channels, culminating in an ultimate sensitivity increase for free $n\rightarrow\bar{n}$ oscillations of three orders of magnitude beyond the previously attained limit obtained at the Institut Laue Langevin (ILL)~\cite{Baldo-Ceolin:1994hzw}. The observation of BNV via free neutron oscillations would be of fundamental significance, with implications for a number of open questions in modern physics which include the origin of the matter-antimatter asymmetry, the possible unification of fundamental forces, and the origin of neutrino mass~\cite{Barbier:2004ez,Mohapatra:2009wp,Calibbi:2016ukt,Babu:2006xc,Babu:2013yca,Phillips:2014fgb}.  

The first stage of this program, HIBEAM (High Intensity Baryon Extraction and Measurement), will employ a fundamental physics beamline during the first phase of the ESS operation.  This stage focuses principally on searches for neutron conversion to sterile neutrons $n'$ which would belong to a ``dark" sector of particles~\cite{Berezhiani:2009ldq,Berezhiani:2017azg,Berezhiani:2020vbe} and which may be observable via mixing i.e. the conversion of long-lived, electrically neutral particles with their dark sector partners and {\it vice  versa}.  The second stage of the program, the NNBAR experiment, is the focus of this paper. 

The NNBAR experiment would take neutrons produced from a moderator which would then be reflected and focused through a magnetic field-free region towards a distant carbon target. The target is surrounded by a detector to observe a baryon number annihilation signal of an antineutron with a nucleon in a carbon nucleus. 

The expected three order of magnitude increase in sensitivity since the last search~\cite{Baldo-Ceolin:1994hzw} rests on the exploitation of a new moderator located below the spallation target (see Section~\ref{moderatorsection}), a field free propagation region, and on improvements in neutron focusing and detector technologies. The design and quantification of the physics potential of the NNBAR experiment forms an important part of the HighNESS~\cite{santoro2020development,Santoro:2022tvi, Santoro:2022qvb} program to study a potential upgrade of the ESS.   
This paper describes the development of a model for the design of the NNBAR experiment. The paper is structured as follows: first, an overview is given of the ESS.
The experiment's sensitivity is dependent on the different components which are then described: the moderator, neutron optics, magnetic shielding and annihilation detector.

%% file: ess.tex
\section{The European Spallation Source}
The European Spallation Source, ESS, when completed at full specifications, will be the world's most powerful facility for research using neutrons ~\cite{Garoby_2017}. The facility's unique capabilities (a  higher useful flux of neutrons than any research reactor, and an unprecedented level of neutron brightness) will exceed those of today's leading neutron sources, enabling new opportunities for researchers across the spectrum of scientific discovery, including materials, life sciences, energy, environmental technology, and fundamental physics. 

There are $15$ instruments currently under construction at ESS~\cite{ANDERSEN2020163402}, representing a subset of the full $22$-instrument suite required for the facility to fully realize its scientific objectives, as defined in the ESS statutes. 
Moreover, the ESS  mandate includes a fundamental physics program and the current lack of a dedicated beamline for fundamental physics has been identified as one of the most important missing capabilities~\cite{ess-gap}.  
In addition to the 15 instruments, a test beamline will be installed among the very first instruments which serve the primary purpose to characterize the target-reflector-moderator system, verifying the performance of the neutron source at the start of the ESS operation.
 
\subsection{Overview of the facility}
\begin{figure}
\begin{center}
\includegraphics[width=.76\textwidth]{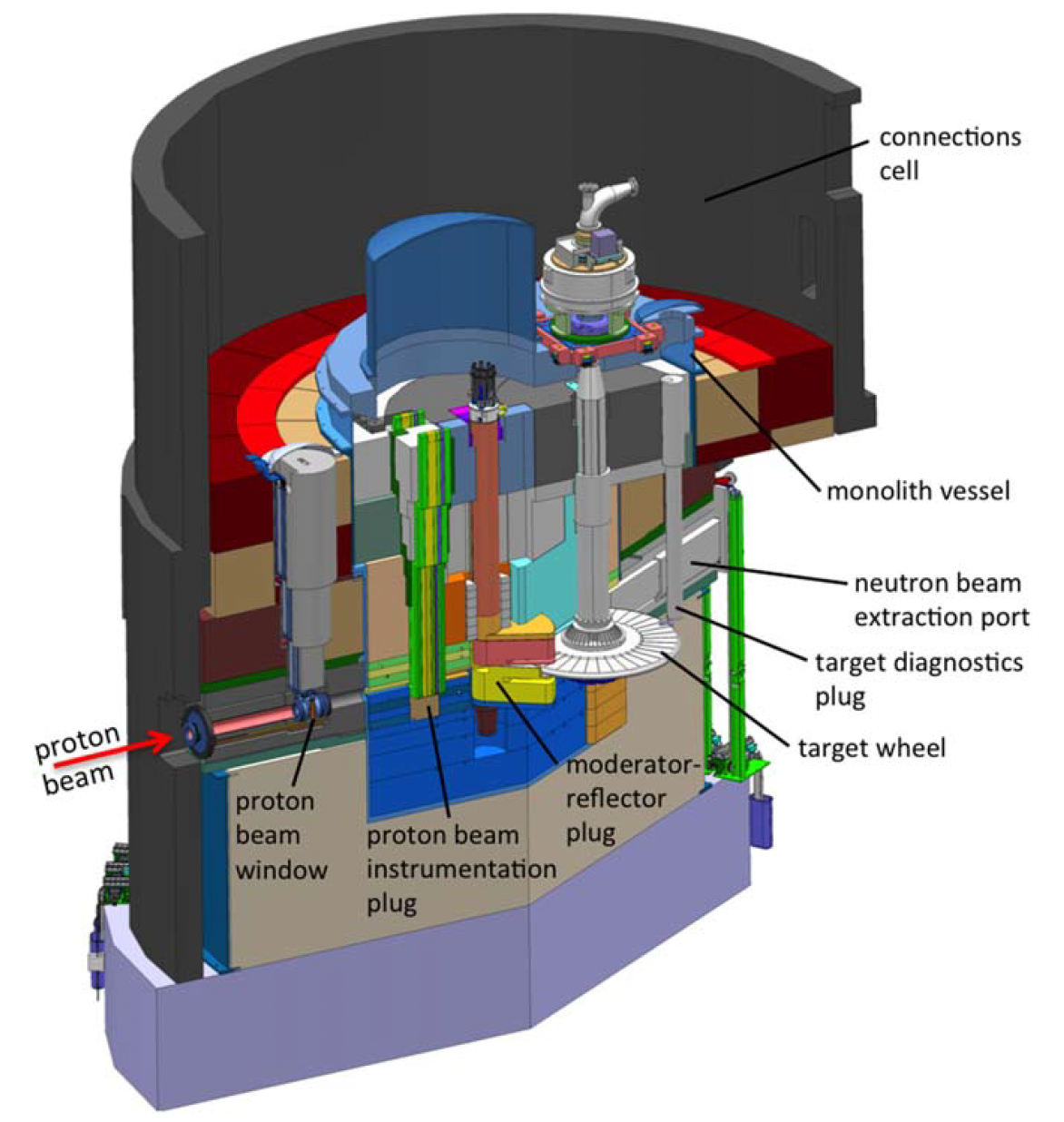}
\caption{A cutaway view of the monolith. The target wheel, the moderator-reflector plug, and other components are shown.}
\label{monolith}
\end{center}
\end{figure}

The ESS uses a linear accelerator to accelerate protons.
The high neutron flux from ESS is  due to the fact that ESS will possess the world's most powerful accelerator and the highest beam power on  target. The proton beam of 62.5 mA is accelerated to 2~GeV with a 14Hz pulse structure (with each pulse  2.86ms long). This will give 5MW\footnote{Note that ESS is currently committed to delivering 2MW as accelerator power. A power of 5MW is part of the upgrade plan (see Section~\ref{timeline}). } average power and a peak power of 125MW. Once the proton beam has reached its final energy, the beam hits a rotating tungsten target to produce neutrons by spallation. These are predominately evaporation neutrons at energies about 2 MeV.  Tungsten blocks mounted on a wheel rotating at 23.3 revolutions per minute successively intercept the proton beam. Pressurized helium gas is used as a cooling fluid, reducing the peak temperature by 150$^{\circ}$ between two shots.
Most of the beam power is dissipated in heat in the target which is located inside a 6000 tons shielding configuration known as the monolith. Figure~\ref{monolith} shows the proton beam, the target, and the monolith structure. 

\begin{figure}
\begin{center}
\includegraphics[width=.86\textwidth]{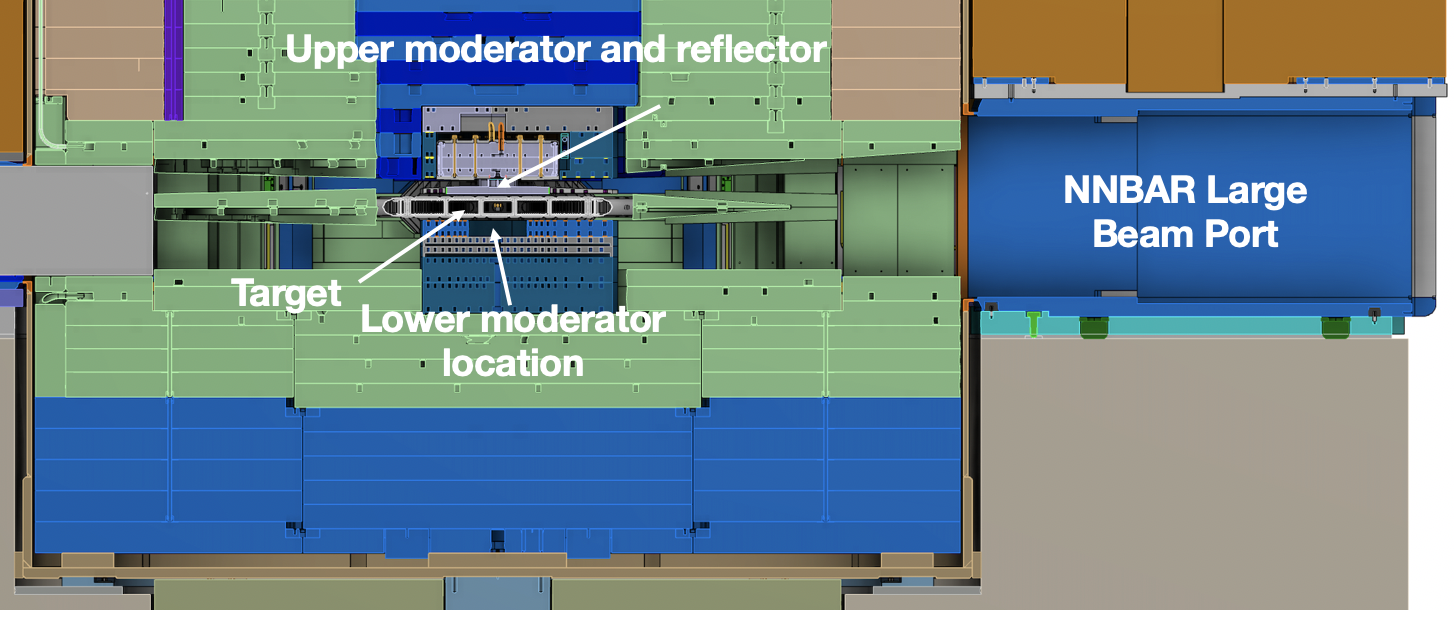}
\caption{Cross sectional view of the ESS target/moderator area and the inner shielding.  In the figure it is shown the location of the ESS upper and lower moderator. The NNBAR experiment will view both  moderators.} 
\label{targetarea}
\end{center}
\end{figure}

The high-energy spallation neutrons are slowed down in the  neutron moderators located inside the moderator-reflector plug and shown in  Figure~\ref{monolith}.  Initially, the ESS will be equipped with only a single compact low-dimensional moderator located above the spallation target\footnote{For this reason we are referring in this paper as the ESS upper ``moderator".} , which has been designed to deliver brightest neutron beams for condensed matter experiments~\cite{zanini_design_2019}, optimized for small samples, flexibility, and parametric studies. A user program will start in 2027, and by 2028 a suite of 15 neutron scattering instruments will be installed making use of the first moderator located above the spallation target. The flexible design of ESS, however, leaves a great opportunity to implement a second source with complementary characteristics going well beyond the initial goals of the facility development.
This source will be installed below the spallation target and for this reason we referred to it as the ``lower moderator", see Figure~\ref{targetarea}.  
The optimization of this source is  part of the NNBAR design. A detailed description of the source is given in Section ~\ref{moderatorsection}.

The target area is surrounded by the monolith, which is a 11-m diameter cylindrical steel structure that also contains the neutron beamports  necessary to extract thermal and cold neutrons from the moderators. The beamport system of ESS is arranged around the moderators and allows the extraction of neutrons above and below the target. This feature will be exploited by the NNBAR experiment that will use neutrons from both the upper and the lower moderators.

\begin{figure}
\begin{center}
\includegraphics[width=.96\textwidth]{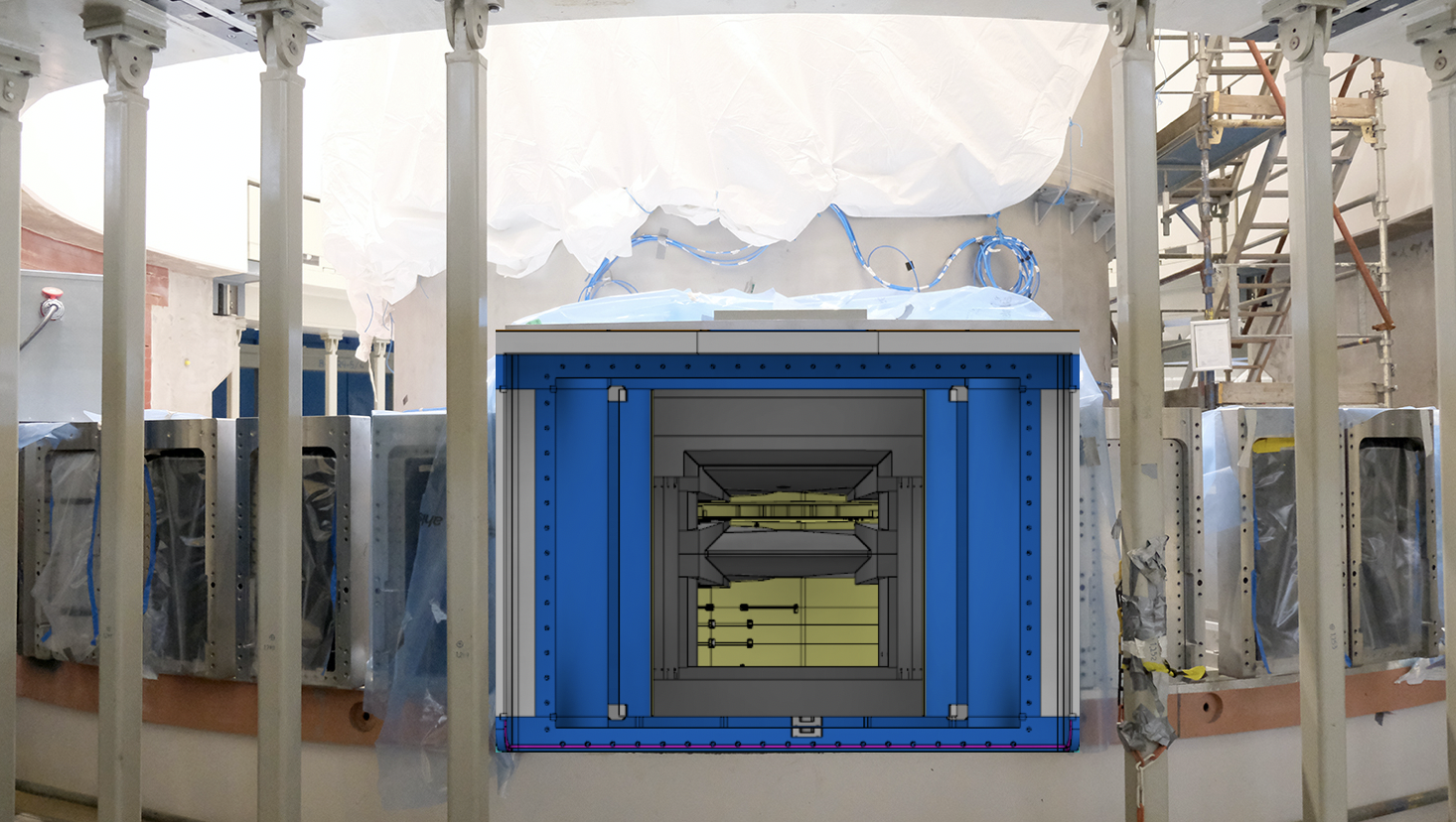}
\caption{Photograph of the  frame of the Large Beam Port being installed in the ESS monolith. A superimposed CAD drawing  is showing the field of view of the LBP. The upper moderator, the inner shielding to avoid a direct view of the target, and the space below the target where the high-intensity moderator will be placed, can be clearly seen.}
\label{lpb}
\end{center}
\end{figure}

The condensed matter instruments currently under construction at ESS will use the standard ESS beamport system while the NNBAR experiment will use a special beamport developed for the experiment, termed the Large Beam Port (LBP) due to its size compared to the standard ESS beamports. The LBP is a critical provision in the ESS monolith, built for the NNBAR experiment to reach its sensitivity goal, that covers a large frame spanning the size of three normally-sized beamports (see Figure~\ref{lpb}). 
Initially, the frame will be filled by three regular-size beamports and at that position, the test beamline will be installed. The three beamports can be removed later to provide the LBP to NNBAR for the duration of the experiment as shown in Figure~\ref{lpb} . At the time of writing, no other existing or planned neutron facilities will have a  beam port of similar dimensions, making the ESS the ideal place for the NNBAR experiment. 

\begin{figure}
\begin{center}
\includegraphics[width=.96\textwidth]{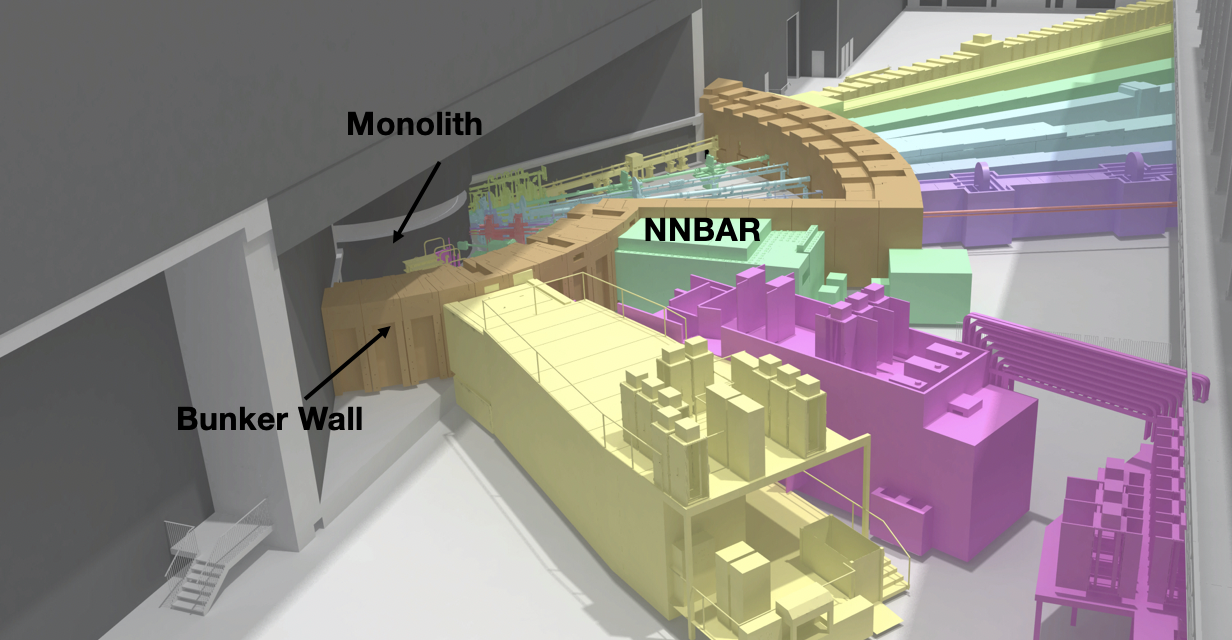}
\caption{The ESS monolith and bunker area. At the start of ESS operation, the NNBAR location will be occupied by the test beamline, used in the early days of the ESS operation to characterize the target-reflector-moderator system. Also shown in the figure are the caves of the LOKI (purple) and FREIA (yellow) instruments. }
\label{bunkerarea}
\end{center}
\end{figure}

After the ESS monolith, the beamlines are housed in the bunker~\cite{bunkerpaper}. The bunker is a common shielding area that surrounds the ESS monolith to protect the instrument area from the high dose of ionizing radiation produced during operation. The shielding structure of the ESS bunker consists of heavy magnetite concrete walls of 3.5m thickness, and a roof, also of heavy concrete, of variable thickness. 
An overview of the ESS bunker is shown in Figure~\ref{bunkerarea} where the future location of the NNBAR experiment is also indicated, initially to be occupied by the test beamline. 

\begin{figure}
\begin{center}
\includegraphics[width=.96\textwidth]{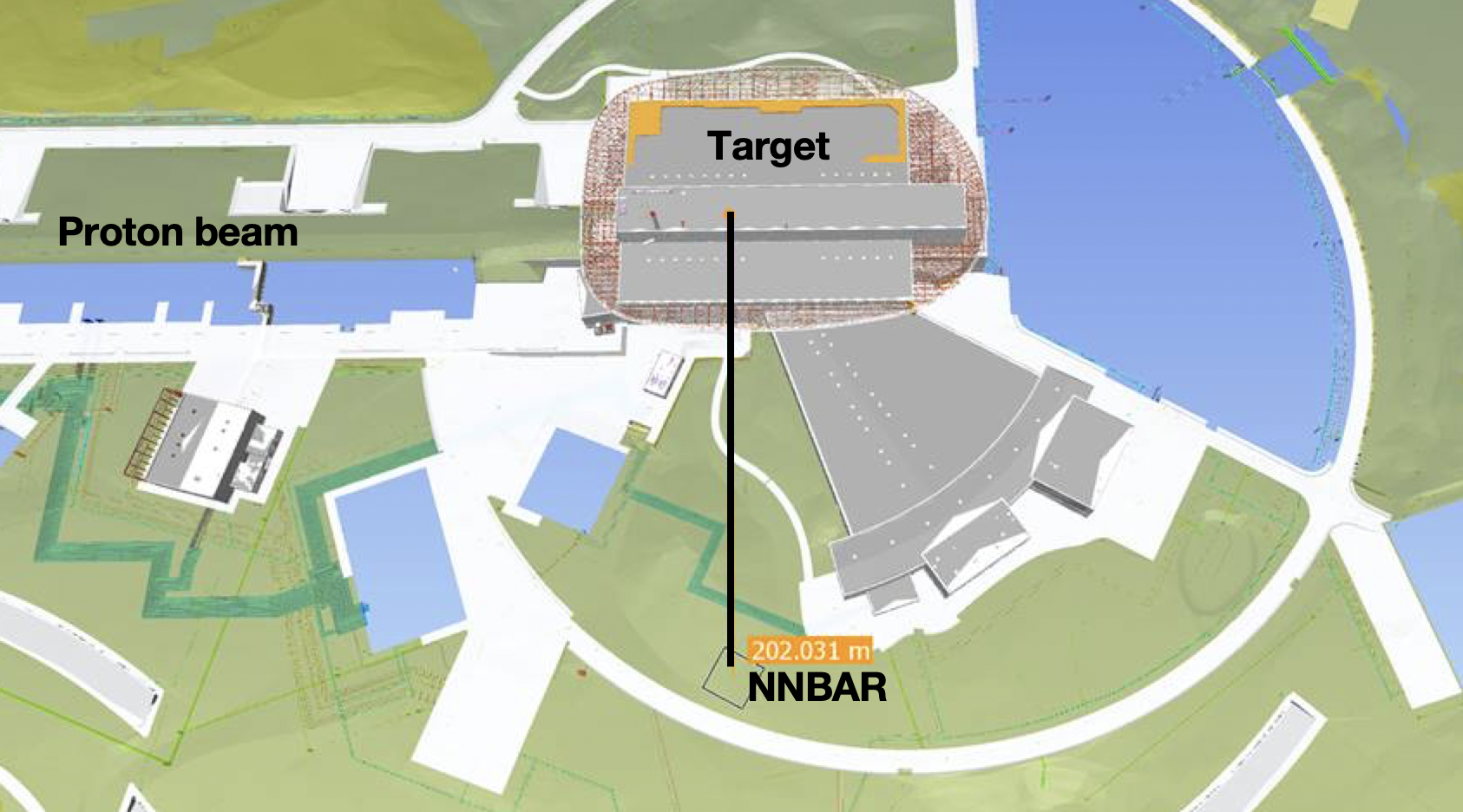}
\caption{NNBAR location in the ESS site. }
\label{essfromthetop}
\end{center}
\end{figure}

As can be seen in Figure~\ref{bunkerarea} and more clearly in Figure~\ref{essfromthetop}, the NNBAR experiment will need an extension of the building since the current instrument hall ends up around 21m from the moderator. This area is available for future upgrades and it is currently a greenfield.

\subsection{ESS timeline and power usage projections}
\label{timeline}

Owing to delays due to the covid-19 pandemic and technical challenges, the ESS has recently conducted a rebaseline exercise to revise its construction and commissioning schedule. The newly revised baseline plan introduces a two year delay and will enable the ESS to start full operations and be open for scientific users working with up to fifteen instruments in late 2027. The maximum accelerator power  and proton beam energies will be 2~MW and 800~MeV, respectively.
The possibility of operating the accelerator at 5MW would then be part of an ESS upgrade project.
The NNBAR experiment is foreseen to take place after the installation of the ESS second moderator which would occur after 2027. Furthermore, the NNBAR experiment will likely not require 5MW power to achieve the planned three orders of magnitude improvement in sensitivity. Studies are ongoing to establish if the lower moderator is able to deliver the performance needed to compensate the lower accelerator power. Further compensation can also be provided by extending the data-taking period. 

%% file: moderator.tex
\section{Design of the NNBAR neutron source}
\label{moderatorsection}
As outlined in Section 2, the configuration of the ESS source offers the possibility of upgrades, including the design of a dedicated moderator for the NNBAR experiment. In fact, so far, only the upper moderator is under construction~\cite{zanini_design_2019} and the space below the spallation target is available for an additional moderator system. The ideal moderator for NNBAR is an intense source of cold neutrons.\\ A preliminary design on a high-intensity voluminous liquid deuterium (L\ce{D_2})  moderator was performed in~\cite{klinkby2014voluminous}, where a gain factor of about 3 in cold neutron beam intensity, with respect to the upper moderator, was calculated. The ongoing design optimization process has the goal to refine and improve the design to further enhance the performance.\\ Here, intensity, or neutron beam current, is defined as the result of integrating the brightness $\Phi(r, E, \Omega, t)$ (angular flux) over the emission surface of the moderator $A$ as given in Equation~\ref{eqt:int}
\begin{equation}
    I_N(E, \Omega_0, t) = \int_A dA \Phi(r, E, \Omega, t)\quad \left[\si{n/s/sr}\right]  
    \label{eqt:int} 
\end{equation}
For the procedure used to calculate the brightness using \software{MCNP}, the reader should refer to the Appendix A.2 in Ref.~\cite{zanini_design_2019}.

\subsection{The model of the ESS moderator}

\begin{figure}[tb]
	\centering
	\includegraphics*[width=\textwidth]{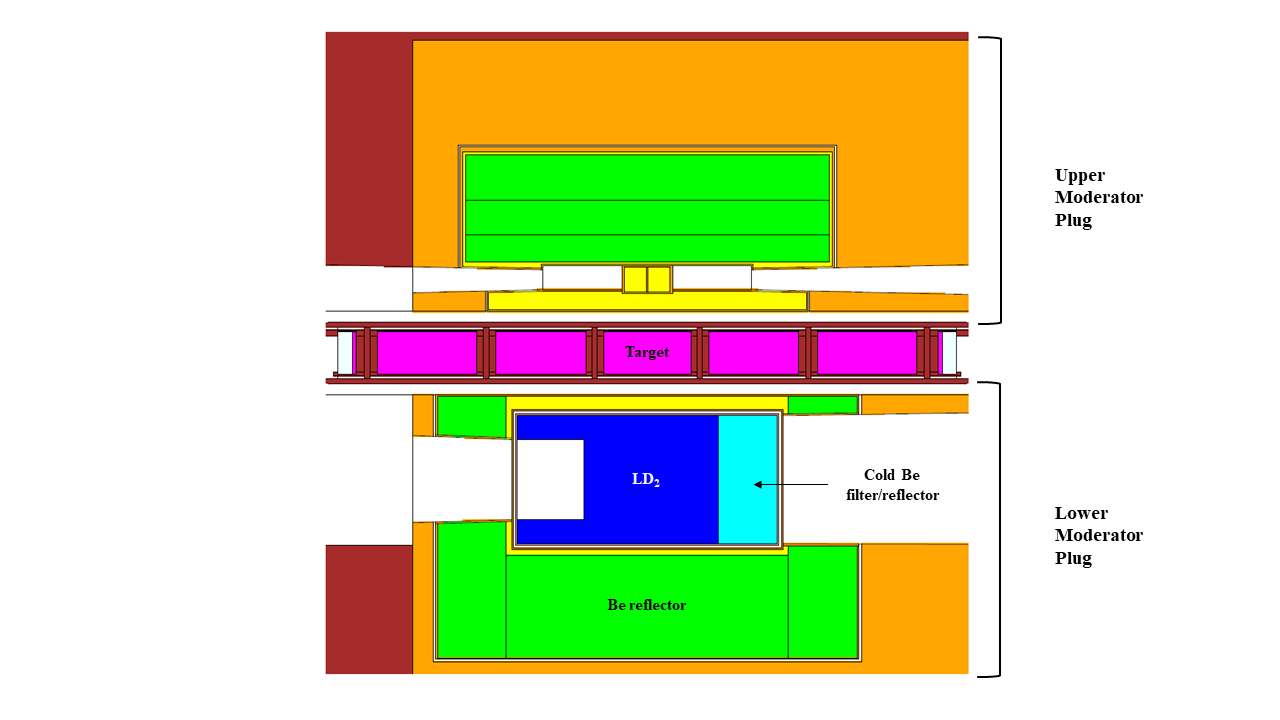}
	\captionsetup{width=0.99\linewidth}
	\caption{\software{MCNP} model of the ESS target with the upper and lower moderator systems. The proton beam is incident on the tungsten spallation target (purple), entering the figure. The NNBAR opening is located on the right side of the moderator while the condensed matter opening is on the left side. Orange and brown = steel (twister frame and inner shielding), green = beryllium reflector, yellow = light water, dark blue = liquid ortho-deuterium, light blue = cold beryllium filter.}
	\label{fig:Mod_MCNP}
\end{figure}

The  source designed within the HighNESS project~\cite{santoro2020development} is intended to serve two classes of applications: fundamental physics, with the NNBAR experiment, and condensed matter science, with the development of  neutron scattering instruments. 
For this reason there are two  emission windows (see Figure \ref{fig:Mod_MCNP}). The orientation of the NNBAR emission window is fixed at 90$^\circ$ to the right of the incoming proton beam, corresponding to the position of the Large Beam Port. Conversely, there are several possibilities for the emission window (or windows) for  neutron scattering instruments, due to the fact that there are several beamports at different angles that could be used. In the design work performed so far we have placed a second opening with size
15 cm x 15 cm at 90$^\circ$ to the left of the incoming proton beam; however, the angle and size of this second opening could change once the position of the neutron scattering instruments using the lower moderator will be chosen. It is also possible that a second opening, most likely of smaller dimensions, could be added for neutron scattering instruments, to add additional flexibility.

After a first round of optimizations the design of the lower moderator 
consisted of a 45 cm  x 48.5 cm x 24 cm box (respectively, along and transverse to the proton beam and vertical dimension) filled with L\ce{D_2} at \SI{20}{K}. The moderator is surrounded by a light water premoderator (\SI{2.5}{cm} thick, facing the target and \SI{1}{cm} in all the other directions) and a water-cooled  beryllium reflector at room temperature. 

The size of the opening in the moderator for the NNBAR experiment is 40 cm x 24 cm, which is the largest surface allowed by the geometry of  the inner shielding  and of the Large Beam Port. This choice follows from the observation that the intensity increases with the size of the emission surface; therefore, the maximum size for the opening was chosen (see Figure \ref{targetarea}). 
Inside the aluminum vessel of the moderator and on the side of the NNBAR emission windows a \SI{11}{cm}-thick beryllium filter/reflector is placed. This block is cooled by the liquid deuterium and it produces an increase of the cold spectrum above \SI{4}{\angstrom}  due to the presence of a Bragg edge in the beryllium scattering cross section at 5 meV. The effect of this filter/reflector is to transmit neutrons with $\lambda > 4~\si{\angstrom}$, and to reflect back faster neutrons, which can have a second chance to be moderated and  exit the surface. A view  of the \software{MCNP6} model is shown in Figure \ref{fig:Mod_MCNP}. 

A parametric study with a single-opening model fixed some design elements such as the ortho-para deuterium concentration, premoderator thickness and moderator temperature. The parameters left to optimize are the moderator dimensions along and transverse to the proton beam, thickness of the Be filter on NNBAR side and depth of the re-entrant hole on the opposite side. The choice of inserting the Be filter is supported by the evidence of enhanced Figure Of Merit (FOM$_{mode}$\footnote{We refer to the Figure Of Merit used for moderator optimization as FOM$_{mode}$.}) for both openings, namely 25\% for NNBAR and 4\% for condensed matter compared with the same case when liquid \ce{D_2} fills the Be filter volume. For NNBAR, the FOM$_{mode}$ is the integrated intensity at the detector for the neutrons between \SI{2.5}{\angstrom} and \SI{15}{\angstrom}, weighted with $\lambda^2$. 

\subsection{Performance of the moderator}
The present design of the  moderator 
gives an intensity of the source, integrated above \SI{4}{\angstrom} of \SI{6.7e15}{n/s/sr}. This is the intensity from the emission window of 24$\times$40 cm$^2$, for 5 MW of beam power, 2 GeV proton energy and 2.5 mA average proton current.   
For 2 MW of beam power (800 MeV proton energy for 2.5 mA average proton beam current), the intensity is simply a factor 2.5 lower, because 
the intensity per unit power is almost identical  between 800 MeV and 2 GeV proton beam energy (see Figure \ref{fig:spectra_MOGA} and discussion below).
 At 2 MW therefore the intensity of the moderator is about \SI{2.7e15}{n/s/sr}. 
It is interesting to note that this number is quite close to \SI{2.9e15}{n/s/sr}, calculated in \cite{klinkby2014voluminous} for 5 MW. That calculation was the basis for the original estimate of the performance increase of NNBAR compared to the ILL measurement \cite{Baldo-Ceolin:1994hzw}. It would therefore mean that the present performance of the NNBAR moderator at 2 MW is essentially equal  to the  performance of the previous design at 5 MW. However, a large part of this gain  is due to the increase in the size of the emission window. It needs to be verified whether the optics system for a large emission window will be as efficient in directing neutrons towards the detector areas as for a smaller emission window. This study is currently on-going (see Section \ref{sec:reflector}) and will be part of the continuation of the design work in an iterative process, aiming to reach the best possible configuration for the experiment.

\subsection{Spectral distribution of neutron intensity}
\begin{figure}[tb]
	\centering
		\includegraphics*[width=0.7\textwidth]{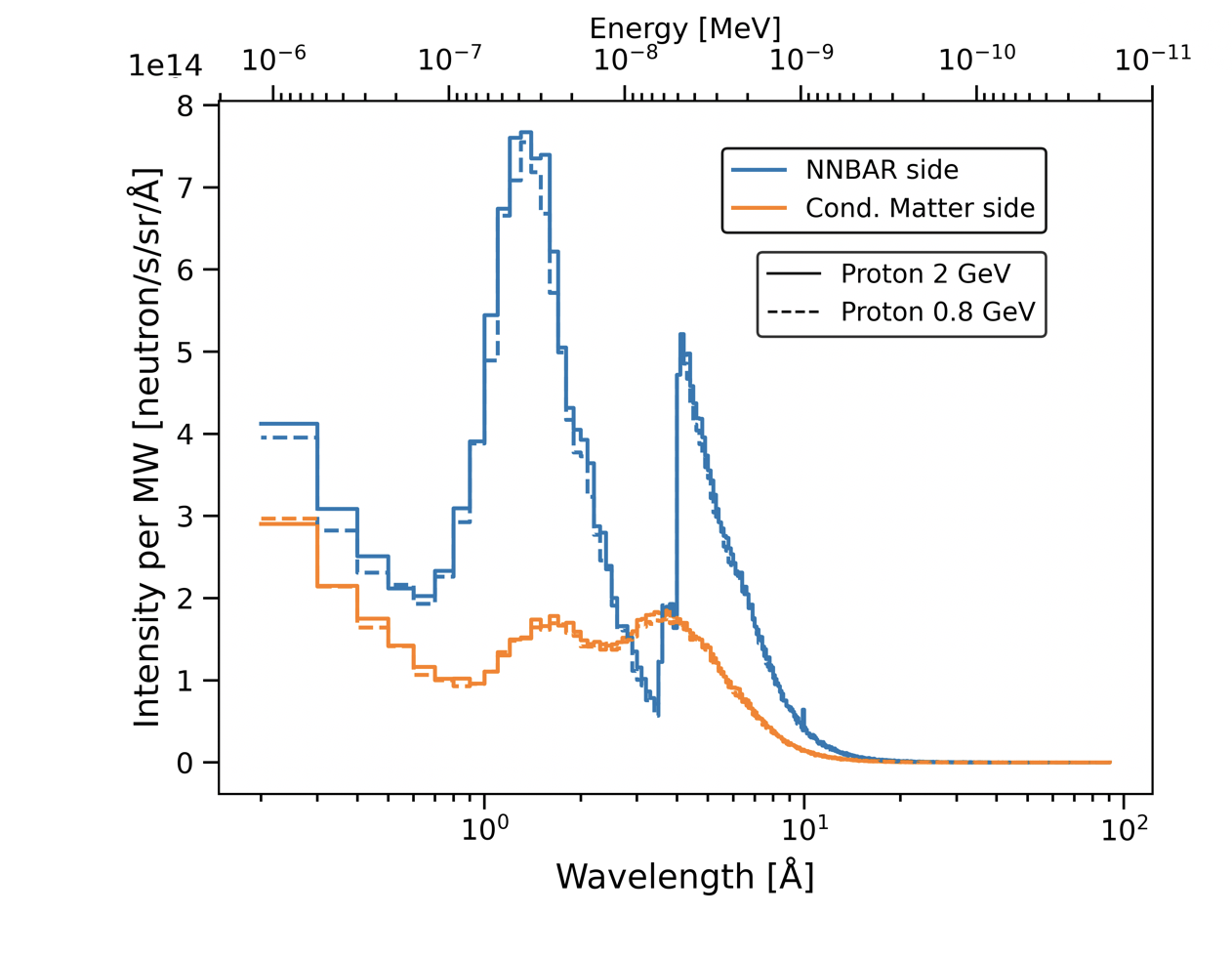}
	\captionsetup{width=0.99\linewidth}
	\caption{ Comparison of the spectra from the two moderator openings for protons impinging on the target  at \SI{2}{GeV} (giving 5 MW of average beam power for 2.5 mA current) and \SI{800}{MeV} (giving 2 MW of average beam power for 2.5 mA current), per unit power of the accelerator.}
	\label{fig:spectra_MOGA}
\end{figure}

The wavelength spectra for the optimized moderator normalized on the accelerator power are shown in \cref{fig:spectra_MOGA}.
The main feature of the wavelength spectrum on the NNBAR side is the  sharp cut-off at \SI{4}{\angstrom} due to the beryllium filter.
The cold spectrum from liquid deuterium is emitted on the other opening, for the neutron scattering instruments. We also compare the spectra per unit power from the two openings when the protons impinging on the target are at \SI{800}{MeV} and \SI{2}{GeV}. The first energy corresponds, to 2.5 mA average current, for the  operation at 2 MW, while the second value corresponds to 5 MW power. 
Although the target and moderator relative positions are optimized for 2 GeV energy, it is interesting that at the location of the NNBAR beamport the spectra per unit power are nearly identical for the two proton energies, indicating that there is no penalty for operating at 800 MeV, besides the obvious drop in absolute intensity due to the lower power.
\subsection{Spatial distribution of neutron intensity}
\begin{figure}[tb]
	\centering
	\includegraphics*[width=\textwidth]{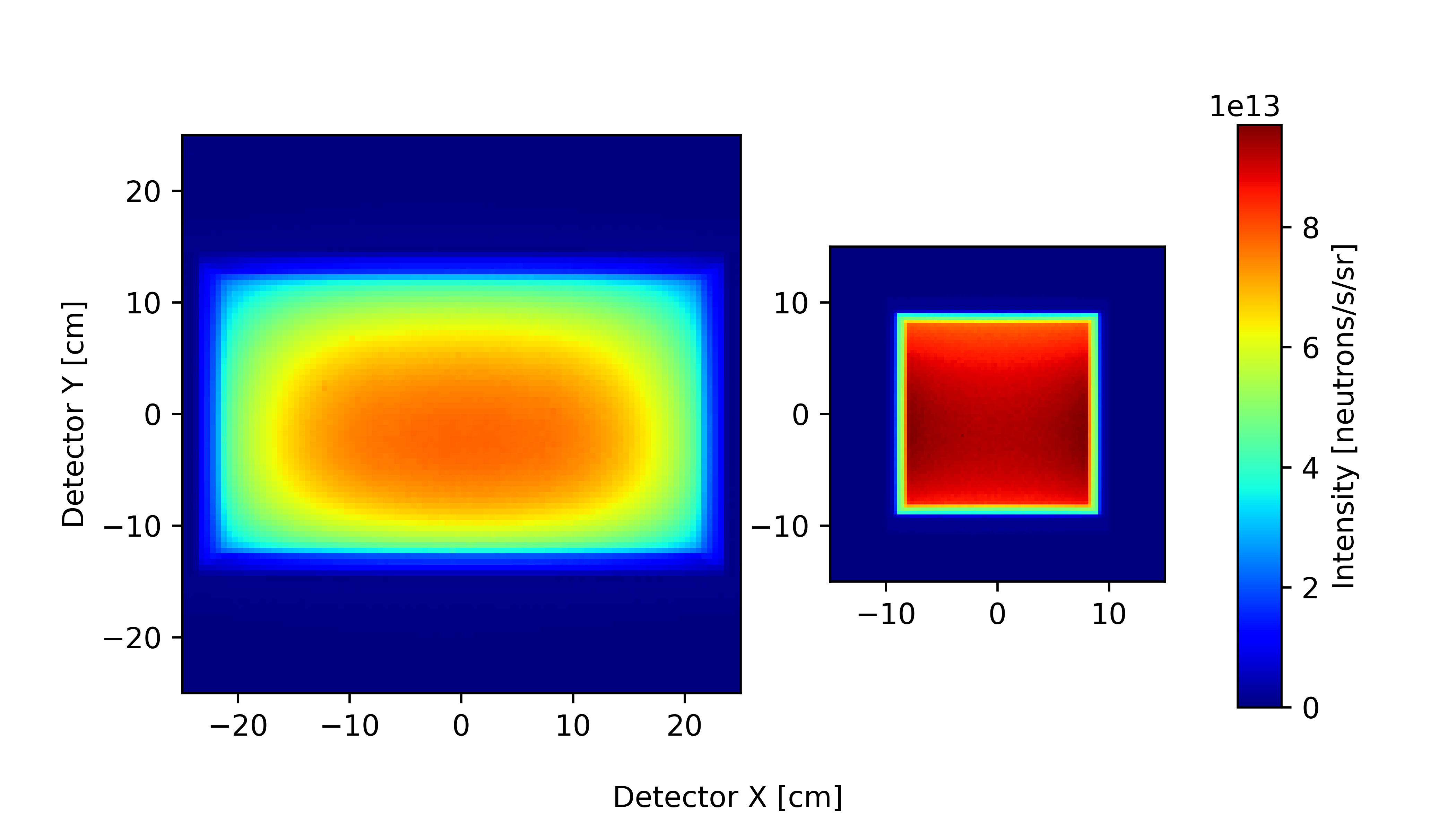}
	\captionsetup{width=0.99\linewidth}
	\caption{Pinhole images of the  24x40 {cm} NNBAR opening (left) and the 15x15 {cm} opening for neutron scattering experiments (right) with a filter for cold neutrons ($\lambda > \SI{4}{\angstrom}$). The pinhole is placed at \SI{2}{m}  from the center of the moderator along the central axis, while the detector is \SI{2}{m} away from the pinhole on the same axis. The pictures  are inverted along the vertical direction due to the camera obscura effect.}
	\label{fig:Pinhole}
\end{figure}

To understand the distribution of the brightness across the moderator surface, a study was made of the spatial distribution of the neutron intensity coming on the surface of the moderator with a pinhole-camera tally integrated with \software{MCNP}. 
The distribution of emitted neutrons across the emission surface is important since it influences the design of the optics to transport the largest amount of neutrons to the detector area. 
In the calculations for this work, a pinhole is placed \SI{2}{m} away from the center of the moderator along the central axis, while the detector is \SI{2}{m} away from the pinhole on the same axis. The result is shown in Figure \ref{fig:Pinhole} for the two openings (NNBAR on the left and for the condensed matter experiments on the right) and $\lambda > \SI{4}{\angstrom}$.
The intensity distribution is fundamentally different between the two openings. The bigger and shallower NNBAR opening has a non-uniform intensity profile along both axes, with a central hot spot slightly off-centered toward the top of the moderator (the pinhole images are inverted along the vertical and horizontal direction). The smaller and deeper opening for the neutron scattering experiments, instead, is a more uniform and brighter source. Even though the integrated intensity is higher on the NNBAR side, the effective gain coming from having such a large opening in the moderator is still under study. This work is being carried on in an iterative way with the focusing reflector team responsible for the neutron transport to the target (see Section \ref{sec:reflector}).

%% file: reflector_shortened.tex
\section{The NNBAR focusing reflector}
\label{sec:reflector}

For NNBAR to utilize the LBP and make the most of the neutron intensity of the ESS source, there must be a reflector to ensure that a large neutron flux from the LBP aperture is directed and focused through a magnetically shielded region onto the annihilation target. The large-aperture neutron reflector will play a major role in the increase of sensitivity of the search. In vacuum and with magnetic shielding, the probability of the $n \rightarrow \bar{n}$ transformation is given by 
$P_{n \rightarrow \bar{n}}=<t^2>/\tau_{osc}^2$,
where $<t^2>$ is the average square of the free flight time of neutrons in the experiment and $\tau_{osc}$ is the characteristic time of $n \rightarrow \bar{n}$ transformation for free neutrons in a vacuum. The current limit on $\tau_{osc}$ was set three decades ago in the state-of-the-art experiment performed with a cold neutron beam at the ILL reactor \cite{Baldo-Ceolin:1994hzw} ($\tau_{osc} \geq 0.86 \times 10^{8} s$). The goal of NNBAR is to increase the sensitivity for the discovery of $n \rightarrow \bar{n}$ processes with free neutrons by a factor of 1,000. The figure of merit of the experiment will be determined by $\mathrm{FOM}_{n\bar{n}} = N \cdot <t^2>$, where $N$ is the number of free neutrons with $<t^2>$ on the detector target. 
%
To have a figure of merit that is easy to compare with previous searches, the $\mathrm{FOM}_{n\bar{n}}$ obtained is normalized (as described in \cite{Addazi:2020nlz}) with respect to the one of the above mentioned experiment at the ILL that ran for one year so that the $\mathrm{FOM}_{n\bar{n}}$ is given as quantity of ILL units per year and a value of $\mathrm{FOM}_{n\bar{n}}=1$ indicates a sensitivity equal to the one achieved at ILL.
All $\mathrm{FOM}_{n\bar{n}}$ values given in this section are in these units and we will be referred in the paper as $\mathrm{FOM}_{n\bar{n}}$.

\subsection{Basic Reflector Setup and Simulations}

\begin{figure}[tbp!] 
	\centering
	\includegraphics[width=1\columnwidth]{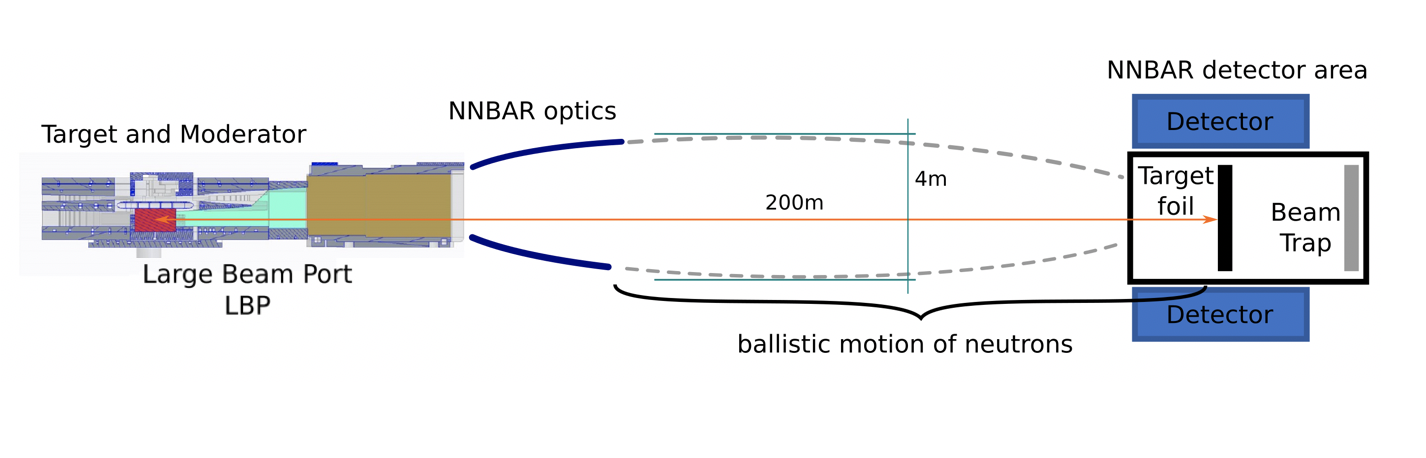}
	\caption{Schematic overview of the NNBAR experiment (not in scale) with a focus on the 
	reflector configuration used in the Monte-Carlo simulations.
	The moderator (source) to detector distance is \SI{200}{\meter}. The transversal dimension of the reflector is bounded by the diameter of \SI{4}{\meter}.
	The annihilation target is of radius \SI{1}{\meter}. The vacuum tube and magnetic shielding are not shown.}
	\label{fig:NNBAR_schematic}
\end{figure}

The principal setup for the reflector that is under study is shown in Figure \ref{fig:NNBAR_schematic}.
Neutrons that are generated at the target are moderated and traverse the large beam port.
A system of elliptical shaped neutron guides is placed in the region behind the LBP's exit to focus the neutrons in the direction of the detector located upstream. The optic is supposed to start at a minimum distance of $\SI{10}{\meter}$ from the moderator center.
After being reflected by the optics, the neutrons fly, free from perturbation, but under gravity force, to the detector region at the end of the instrument. The moderator (source) to detector distance is foreseen to be around $\SI{200}{\meter}$. The flight time is measured from the point in time of the last interaction (reflection) with the optic.   The transversal dimension of the reflector is bounded by a maximum assumed tube diameter of $\SI{4}{\meter}$.

To compare different geometries as well as the exact placement of the reflector system, neutron ray-tracing simulations are performed using \software{McStas} \cite{willendrup2020,willendrup2004_McStasNew}. 
%
%
%
The input sources for the simulations are \software{MCPL}-files that have been created from \software{MCNP} simulations of the moderator design (see also chapter \ref{moderatorsection}).
%
%
%
The virtual detector of the simulation is of size 
$\SI{10}{\meter}$ x $\SI{10}{\meter}$ 
and hence distinctively larger then the real NNBAR detector (see Section \ref{nnbardetector}). The $\mathrm{FOM}_{n\bar{n}}$ is calculated for the area of a circle with \SI{1}{meter} radius and centered at the position where the sensitivity is at maximum. 
By running different reflector designs with several different geometries and adjustment of various reflector parameters (e.g. starting position, length, etc.) a large number of such optics can be investigated to find the optimal parameters.
%
%
%
%

%

\subsection{Baseline design}
\label{baseline}
\begin{figure}[tb]
	\centering
    \includegraphics*[width=0.42\textwidth]{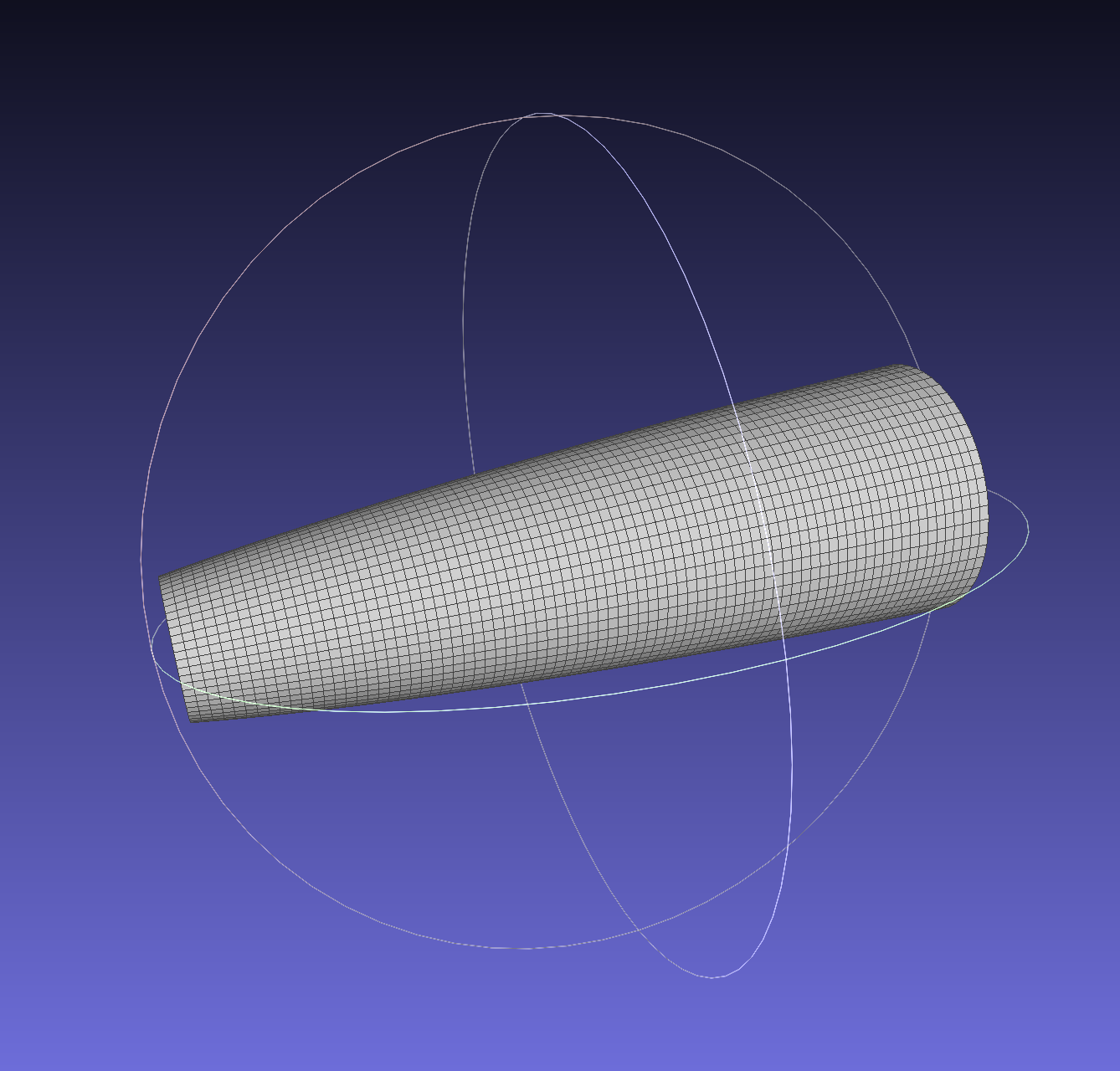}
	\includegraphics*[width=0.56\textwidth]{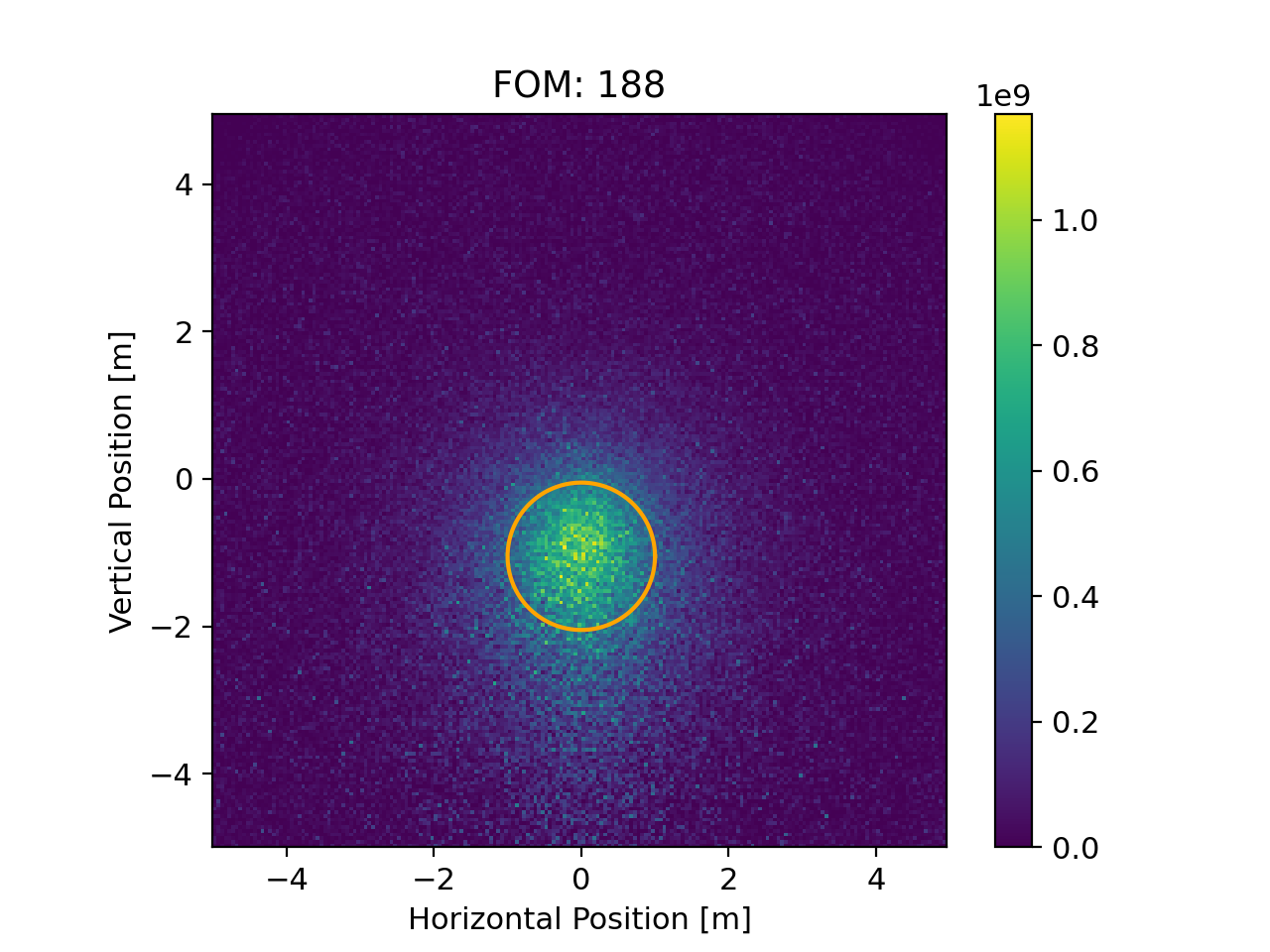}
	\captionsetup{width=0.99\linewidth}
	\caption{Left: 3D visualization of the \SI{40}{\meter} long baseline reflector (axis are not in scale). Right: Result of a \software{McStas} simulation with the baseline reflector and  \SI{2}{\mega\watt} target power. The orange circle marks the detector area of \SI{1}{\meter} radius. A $\mathrm{FOM}_{n\bar{n}}$ of 188 was calculated for target power of 2MW, while for  \SI{5}{\mega\watt} the $\mathrm{FOM}_{n\bar{n}}$ is 470.}
	\label{fig:basline_reflector}
\end{figure}

A possible design (called "baseline" for comparison purposes)  to transport neutrons that diverge from a source to a detector is an elliptically shaped guide made of reflecting material, 
where the source and the detector are placed in the two focal points of the ellipse. 
An elliptical mirror has the property to reflect a ray that originates in the first focal point into the direction of the second focal point.
The baseline ellipsoid is defined by a distance of $\SI{200}{\meter}$ between the two foci 
and a small semi axis $b$ of $\SI{2}{\meter}$. The center of the source (moderator) is located in one focus, while the center of the detector is located in the other focal point.
The reflector covers the part of the ellipse that starts at 
$\SI{10}{\meter}$ from the source and ends at a distance of $\SI{50}{\meter}$ and is therefore $\SI{40}{\meter}$ long.
%
This baseline design scheme was previously used for optimization of parameters and for comparison of several NNBAR configurations in previous publications \cite{Addazi:2020nlz,Frost:2019tom, Phillips:2014fgb}.
%
%
%
A \software{McStas} simulation performed with this reflector with the horizontal axis at the center of LBP gave a $\mathrm{FOM}_{n\bar{n}} = 188\;(470)$ with the moderator parameters and spectrum shown in the previous section for ESS target power of \SI{2}{\mega\watt} (\SI{5}{\mega\watt}).
The shape of the reflector and the focused beam distribution at the annihilation target obtained in the simulation are depicted in Figure~\ref{fig:basline_reflector}.


\subsection{Other reflector geometries options}
\subsubsection{Differential reflector}
\begin{figure}[tbp!] 
	\centering
	\includegraphics[width=1\columnwidth]{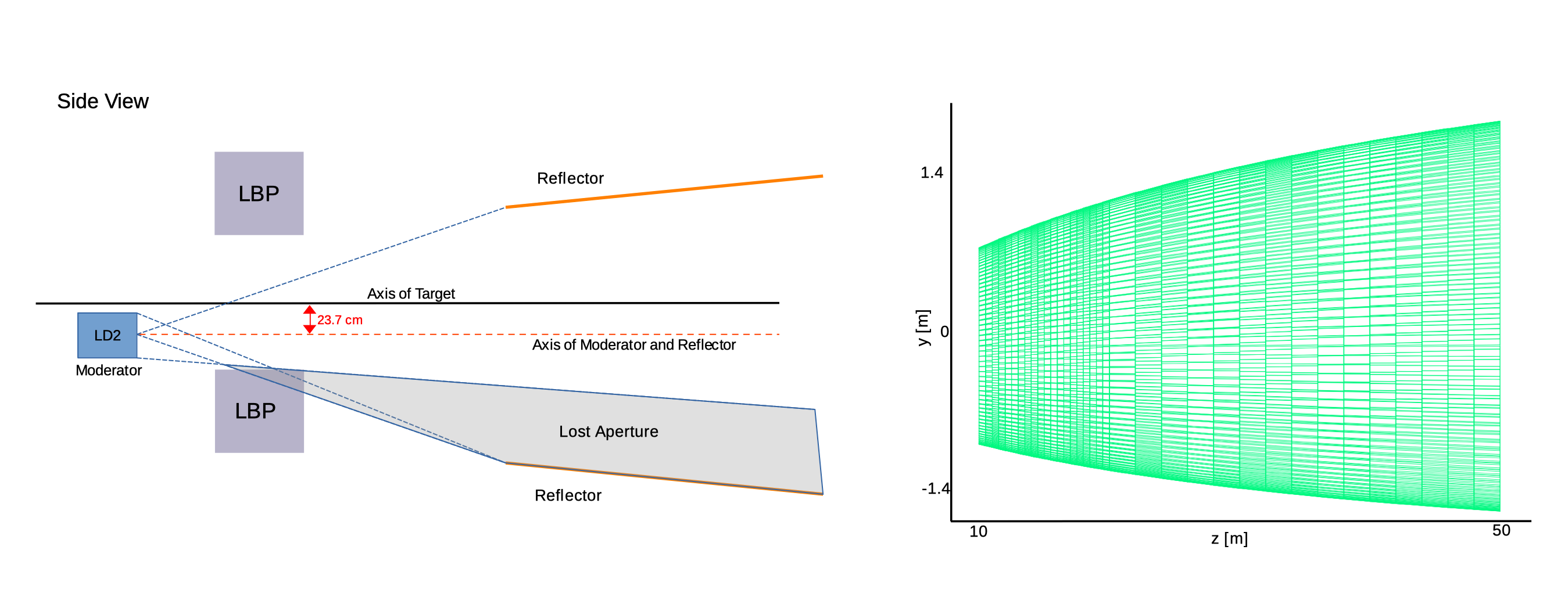}
	\caption{Left: Possible loss of the LBP aperture with horizontal beam axis determined by the position of cold neutron source. This loss is the main argument for considering the option of a ``differential reflector''.
	Right: Depiction of the ``differential reflector''.}
	\label{fig:overshoot_and_diffreflector}
\end{figure}

The center of lower moderator is $\SI{23.7}{\cm}$ below the axis of the LBP (see Figures \ref{targetarea} and \ref{fig:Mod_MCNP}), hence the full aperture of the LBP from the lower moderator is tilted up by $\approx  \SI{2}{\degree}$. 
If the axis of the NNBAR beam, as determined by the reflector position, is horizontal, it would lead to the loss of aperture as explained in the Figure \ref{fig:overshoot_and_diffreflector}. 
%
To cope with this issue the concept of a ``differential reflector'' was proposed \cite{Frost:2019tom}. 
The reflector is positioned exactly in the middle of the LBP and has a distorted ellipsoid shape (see right side of Figure \ref{fig:overshoot_and_diffreflector}). The constituting panels fulfill the solution of a coupled differential equation, to behave on each position like an elliptical mirror and form a continuous surface. 
This reflector focuses and bends the neutron beam by a few degrees in the vertical direction at the same time. This will allow the preservation of the $\mathrm{FOM}_{n\bar{n}}$ with the horizontal beam axis between the centers of the cold source and the annihilation detector. 
%
%
%
\software{McStas} simulations performed with the ``differential reflector'' design support this claim. They gave a $\mathrm{FOM}_{n\bar{n}} = 225$ for \SI{2}{\mega\watt} target power and 562 for \SI{5}{\mega\watt} target power, which amounts to an increase of \SI{20}{\percent} compared to the baseline.
%
\subsubsection{Nested optics}
\label{nestedpar}
\begin{figure}[t]
	\centering
	\includegraphics[width=.8\textwidth]{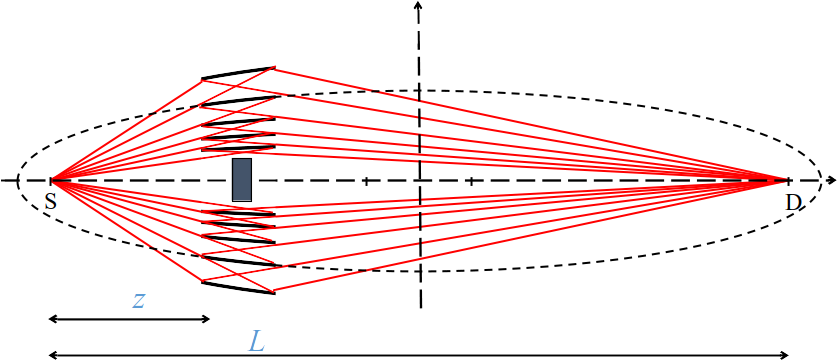}
	\caption{Schematic of a nested elliptical guide. $S$ and $D$ are the common focal points of the ellipses, separated by the distance $L$, form the layers (the dashed line shows one of them).}
	\label{fig:conctruction-nested_optics}
\end{figure}

\begin{figure}[htbp!]
	\centering
	\includegraphics[width=1\textwidth]{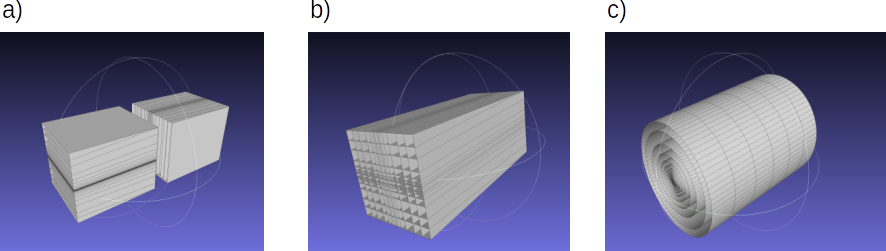}
	\caption{Types of nested optical components a) mono planar b) double planar c) cylindrical.}
	\label{fig:types_of_optics}
\end{figure}

Both the reflector in its baseline configuration and the ``differential reflector'' will be large, mechanically complicated, logistically difficult to assemble and install, and therefore expensive. Different options are therefore being considered, in particular, ``nested reflectors'', 
as proposed in \cite{zimmer2016_MultimirrorImaging, zimmer2019_ImagingNestedmirror} and depicted in Figure \ref{fig:conctruction-nested_optics}.
These reflectors are easier to assemble, they can be much shorter in $\Delta{z}$ along the beam axis and, after optimization, may represent a more economical solution for the NNBAR experiment. 
%
%
%
If the outer layer of such a nested elliptical guide is given, the inner layers can be constructed in a recursive manner such that the layers will not shadow themselves. 
%
%
Different nested layouts of the reflector that are symmetrical around the $z$-axis are possible, namely: (a) a mono planar, (b) a double planar, and (c) a cylindrical system. In figure \ref{fig:types_of_optics} the cross sections of these different types are depicted.  
%
%
%
A difficulty of the nested reflector design is the thickness of the glass substrate used for the construction of the stable high-quality industrial super-mirrors. Recent developments of self-sustaining substrateless super mirrors \cite{Subless} might be an elegant solution to this problem. 
%
%

%

\subsection{Reflector simulation results}
\begin{figure}[htbp!]
	\centering
	\includegraphics[width=1\textwidth]{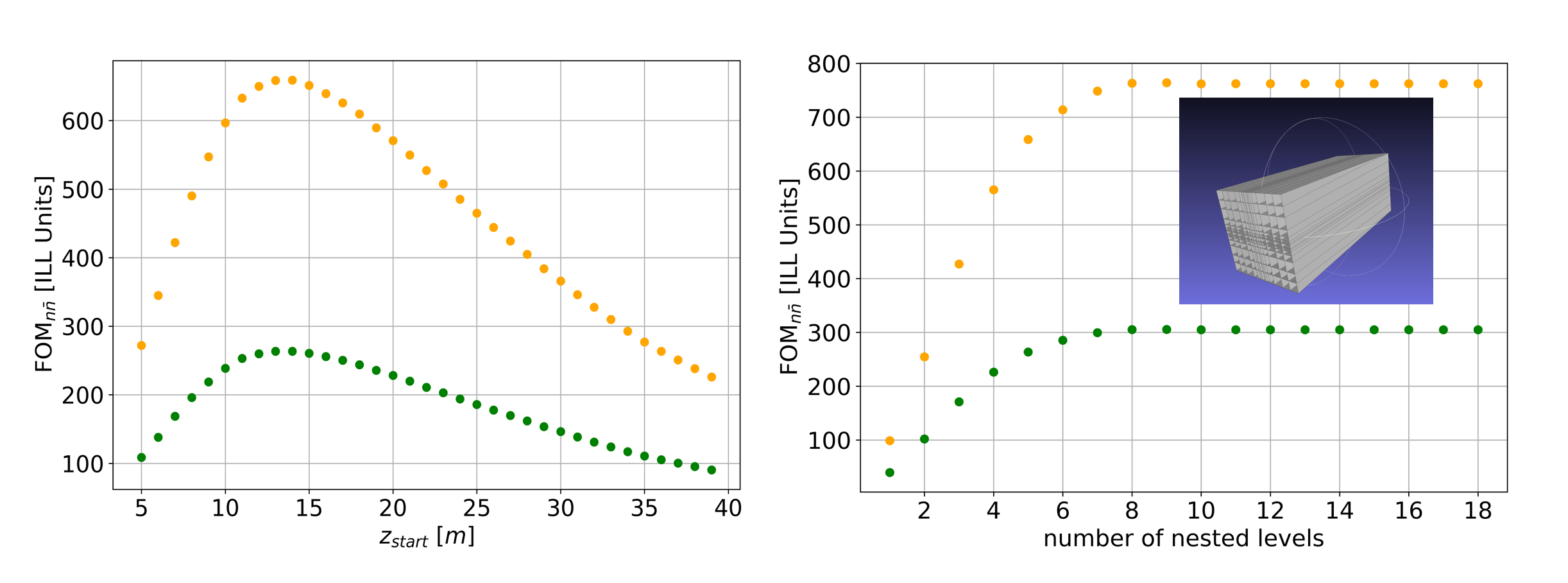}
	\caption{Result of simulations for a nested double planar reflector of length \SI{10}{\meter} (see inlay) for \SI{2}{\mega\watt} (green) and \SI{5}{\mega\watt} (orange) target power.   Left: Varying the starting point of the optic. Right: Increasing the number of nested levels.} 
	\label{fig:_reflector_example_scan}
\end{figure}

The optimal parameters for the different possible geometries are studied by performing various simulations and comparing the obtained $\mathrm{FOM}_{n\bar{n}}$s.
Two such scans for a \SI{10}{\meter} long double planar nested reflector are shown in Figure \ref{fig:_reflector_example_scan}. The scan for the $z_{start}$ parameter (the start of the optic defined as the distance from the moderator) shows an optimum at about \SI{14}{\meter}.
For the number of nested levels, one observes a saturation after a certain point, when adding further levels does not lead to a further increase of the $\mathrm{FOM}_{n\bar{n}}$. 
In the following Figure \ref{fig:collected_results} the results of several simulations with various reflector geometries are summarized. 
Significantly higher $\mathrm{FOM}_{n\bar{n}}$s 
as with the baseline can be achieved with short nested reflectors of the elliptical and double planar type. 
The gains are at least 53\% over the baseline reflector and at least 28\% over the differential reflector. 

\begin{figure}[htbp!]
	\centering
	\includegraphics[width=0.75\textwidth]{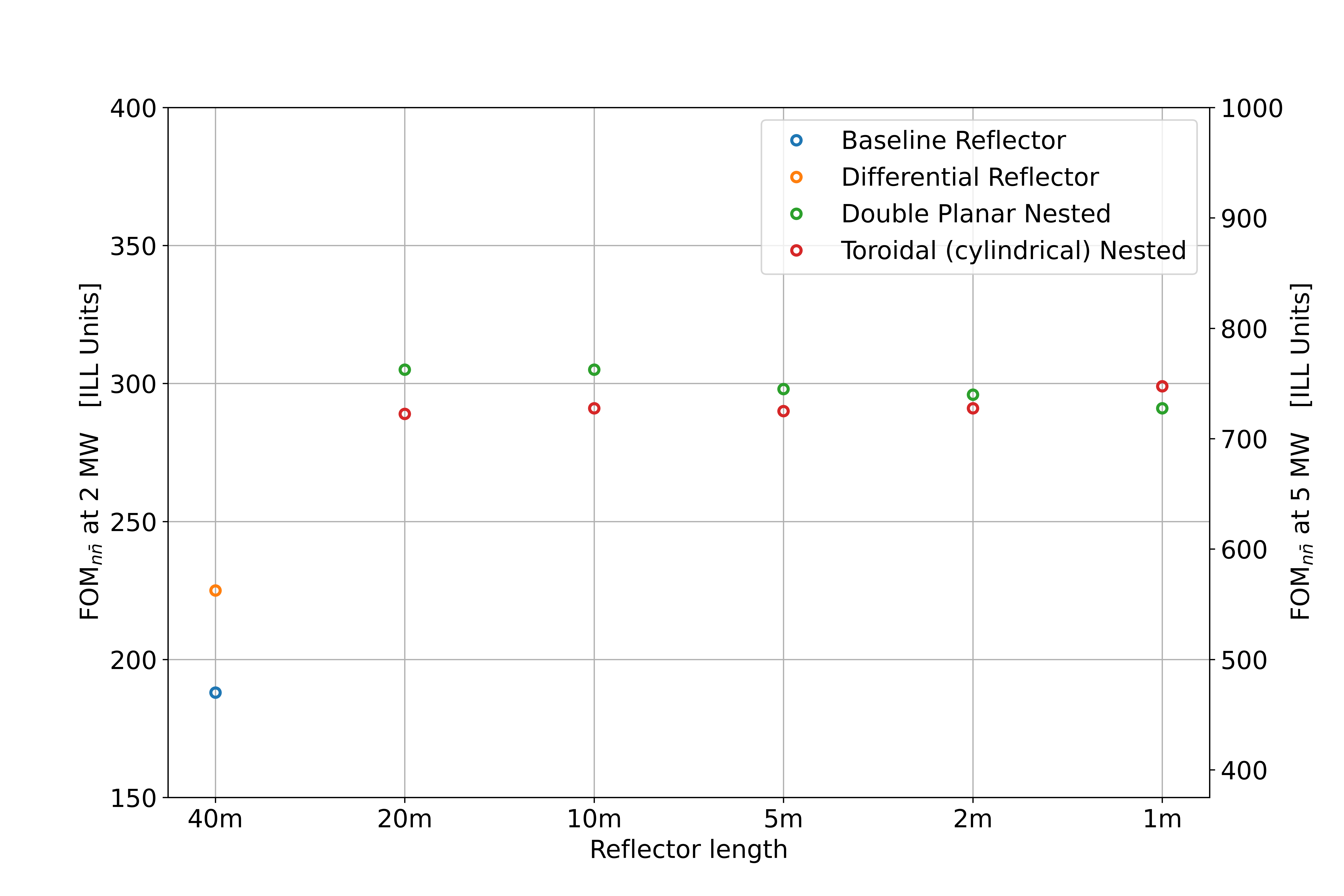}
	\caption{Collected $\mathrm{FOM}_{n\bar{n}}$s for different reflector geometries. (Left axis) Target power of \SI{2}{\mega\watt}. (Right axis) Target power of \SI{5}{\mega\watt}.} 
	\label{fig:collected_results}
\end{figure}

The ESS pulse shape has not yet been taken into account with the concepts presented in this work. An idea of a time-varying focusing reflector to compensate effects of gravity that is driven by piezo-actuators has been discussed previously in \cite{piezo} and might be part of future studies.

%% file: magnetics.tex
\section{Magnetic shielding and vacuum}

Magnetic fields must in principle be small enough to not suppress the transition from the neutron to the antineutron state, which do not have the same energy in a magnetic field due to their magnetic moments. 
%
%
Suppressing the magnetic field results in the so-called quasi-free condition $\|E\|t \ll 1$~\cite{Phillips:2014fgb}, where $E$ and $t$ are neutron energy and propagation time, respectively. 
This condition is fulfilled when the magnetic field along the flight path is less than approximately 10~nT~\cite{Davis:2016uyk} . 
Neutrons flying 200 m inside a shielded environment will however experience a time-varying field in the frame of the particle, which results from gradients, spatial and temporal distortions. 
It has been shown in Ref.~\cite{Davis:2016uyk} that such variations do not significantly suppress the neutron-antineutron conversion, with the effect being smaller for higher neutrons velocities.
Instead, the average field magnitude is critical. 
The aim for NNBAR is thus for neutron propagation in a magnetic field of 5~nT, assuming local changes of similar magnitude and an average field along each neutron trajectory of below 10~nT. A further requirement is that, to avoid interactions, the neutrons propagate in a vacuum. This section describes the design of the magnetic shield and the vacuum system for the experiment.

\subsection{Magnetic shielding}

\begin{figure}
\centering
    \includegraphics[width=\hsize]{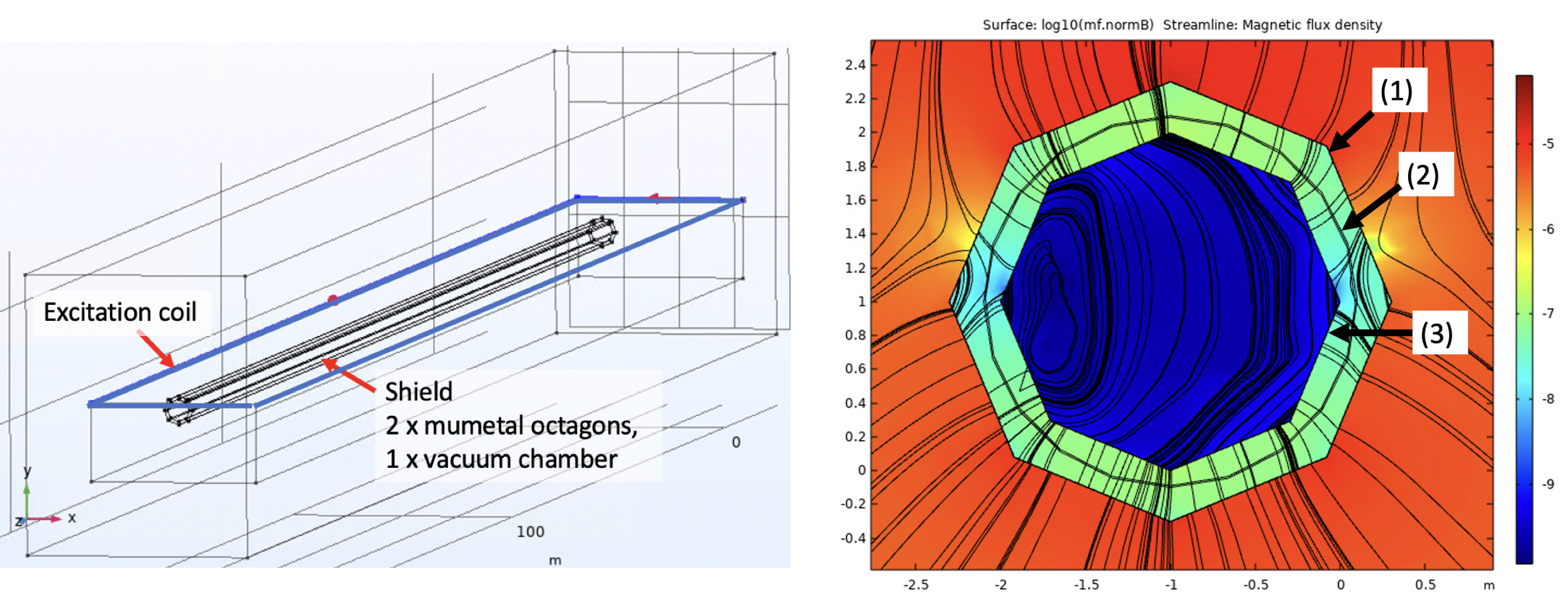}
    \caption{Magnetic shield concept, taken from a \software{COMSOL}~\cite{multiphysics1998introduction} simulation. Left: The 200 m long and 2 m inside diameter arrangement consists of 2 octagons made from mu-metal, with 200 m length and a stainless-steel vacuum chamber arranged in between; an external excitation coil is placed around the shield to calculate the performance. Right: cut through the shield, with log-scale magnetic field magnitude. The asymmetric deformation is due to the orientation of the earth's field. (1) Outer mu-metal shield; (2) vacuum chamber; (3) inner mu-metal shield.}
    \label{fig:magconcept1}
\end{figure}

The concept of the magnetic shield for the NNBAR experiment is shown in Figure~\ref{fig:magconcept1}. It consists of a two-layer octagonal mu-metal shield, combined with a metallic (316L stainless steel) vacuum chamber, which also contributes to the shielding.

Mu-metal shielding provides a static reduction of the magnetic field as well as damping of external magnetic field changes up to about 10 Hz. 
Shielding of higher frequencies relies on the conductivity of the vacuum chamber in combination with the mu-metal. 
The arrangement of the mu-metal is based on a proven small-scale design of the magnetic shield of an atomic fountain \cite{wodey2020scalable}, which consists of mu-metal sheets, which are arranged in an octagonal shape and clamped together.
Overlaps of 50~mm width of the mu-metal sheets are foreseen to ensure proper magnetic flux connection while being a reasonable compromise with magnetic equilibration.
Through this approach, the shield parts can be independently assembled and also detached whenever needed. 
The vacuum chamber can be arranged in between the mu-metal layers or inside the inner mu-metal layer. 

To obtain a low magnetic field, the mu-metal must be magnetically equilibrated~\cite{altarev2015minimizing} through a set of coils, which surround each shield layer independently as a toroidal coil. 
The diameter of the shield is determined by the volume accessible to the neutrons and an approximately 20~cm distance to the shield walls, where the fields after equilibration are too high for the measurement. 
Also, the gap between the mu-metal shells is determined from simulations and set to a minimum of 20~cm.
No details for the section close to the neutron source are yet included in the design.
\subsubsection{Simulations}
\begin{figure}
\centering
    \includegraphics[width=.66\textwidth]{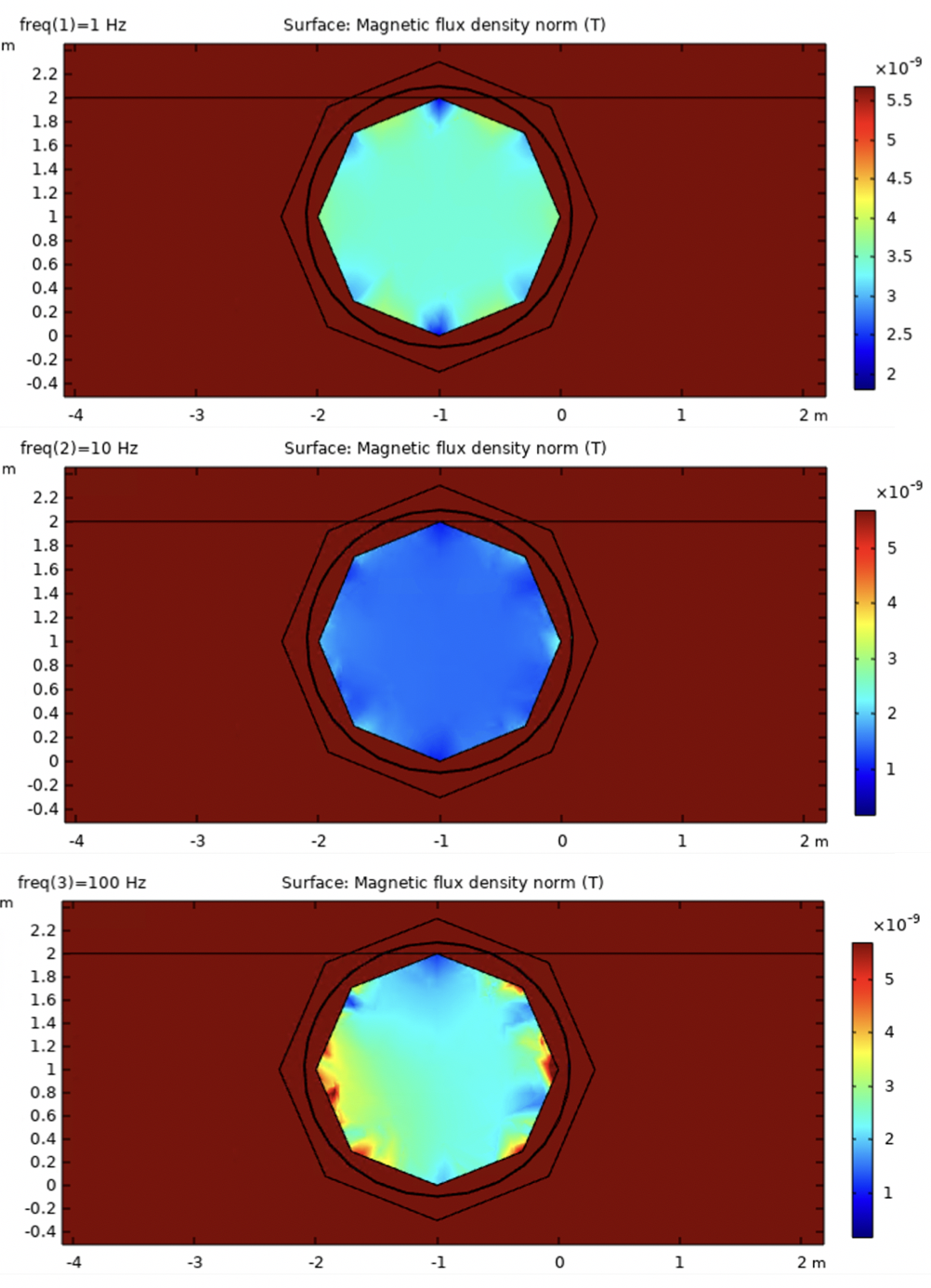}
    \caption{Field inside the shield (cross section in middle region), for 1, 10 and 100 Hz external excitation. }
 \label{fig:onetohundred}
\end{figure}

A scheme of the shield as used for simulation is shown in Figure~\ref{fig:magconcept1}. 
The simulations are based on finite element calculations using the commercial software \software{COMSOL}~\cite{multiphysics1998introduction}. 
For the optimization of the parameters, the shield has been placed inside a static background field normal to the shield axis, different static fields from magnets from the ESS instruments  LOKI, ESTIA, SKADI, DREAM, HEIMDAL, and T-REX~\cite{ANDERSEN2020163402} have also been considered, as well as exemplary tests of steel rebar of magnetic concrete and structures.
A large external coil is placed surrounding the shield, generating a magnetic field perpendicular to the shield axis.
The coil is placed closer to one open end of the shield than on the opposite side to also illustrate the effect of fields entering the shield longitudinally.
Shielding efficiency is then calculated for 1, 10, and 100 Hz, including permeability and currents. Calculation of Direct Current (DC) field reduction is not quantitatively possible for very low fields with magnetostatic means. 
However, the effect of magnetic equilibration is well tested and can be calculated in specific (simple) cases \cite{sun2016dynamic,sun2020limits} and can thus be scaled from a magnetostatic simulation based on a typical mu-metal hysteric curve.
Figure \ref{fig:onetohundred} shows the fields inside the shield from excitations with 1 to 100 Hz, with an external excitation with the same amplitude.

The SF at 0.01 Hz, determined from a simulation at 1 Hz and scaled using experimental data, is $\sim$10-300 along the central axis, depending on the position in the shield. 
At 50~Hz, the SF is $>$~1000, including the vacuum chamber conductivity.
After variation and minimization of thicknesses of inner and outer mu-metal shield layers, the DC field is expected to be $<$~5~nT after degaussing for 1.5~mm layer thickness for both shields.

\begin{figure}
\centering
    \includegraphics[width=.56\textwidth]{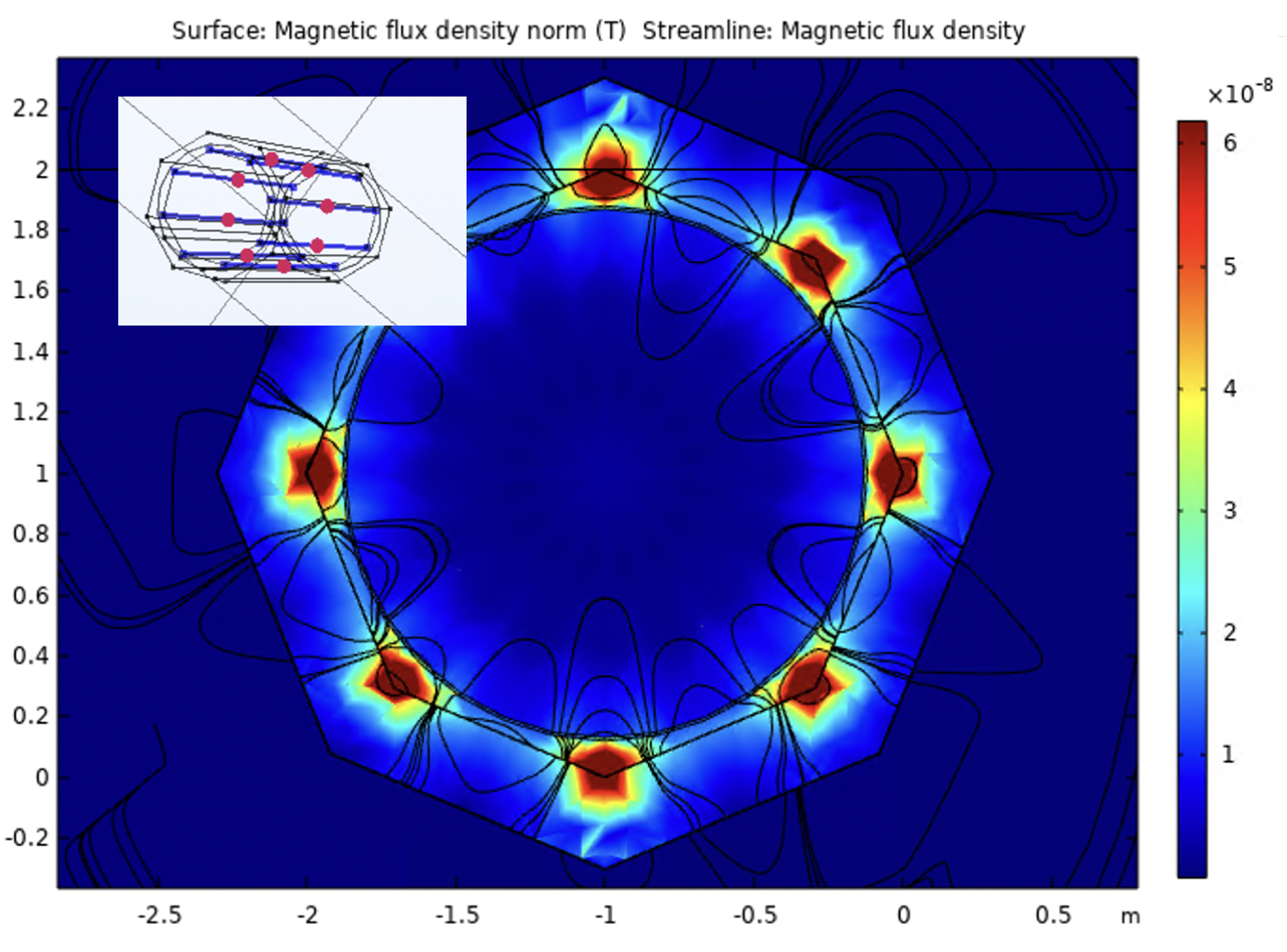}
    \caption{Field pattern expected as result of the magnetic equilibration process. }
 \label{fig:fieldinside}
\end{figure}

Although the degaussing efficiency is not simulated quantitatively but scaled from experimental findings, the field distribution inside the shield after degaussing is modeled, see Figure~\ref{fig:fieldinside}.

\begin{figure}
\centering
    \includegraphics[width=\hsize]{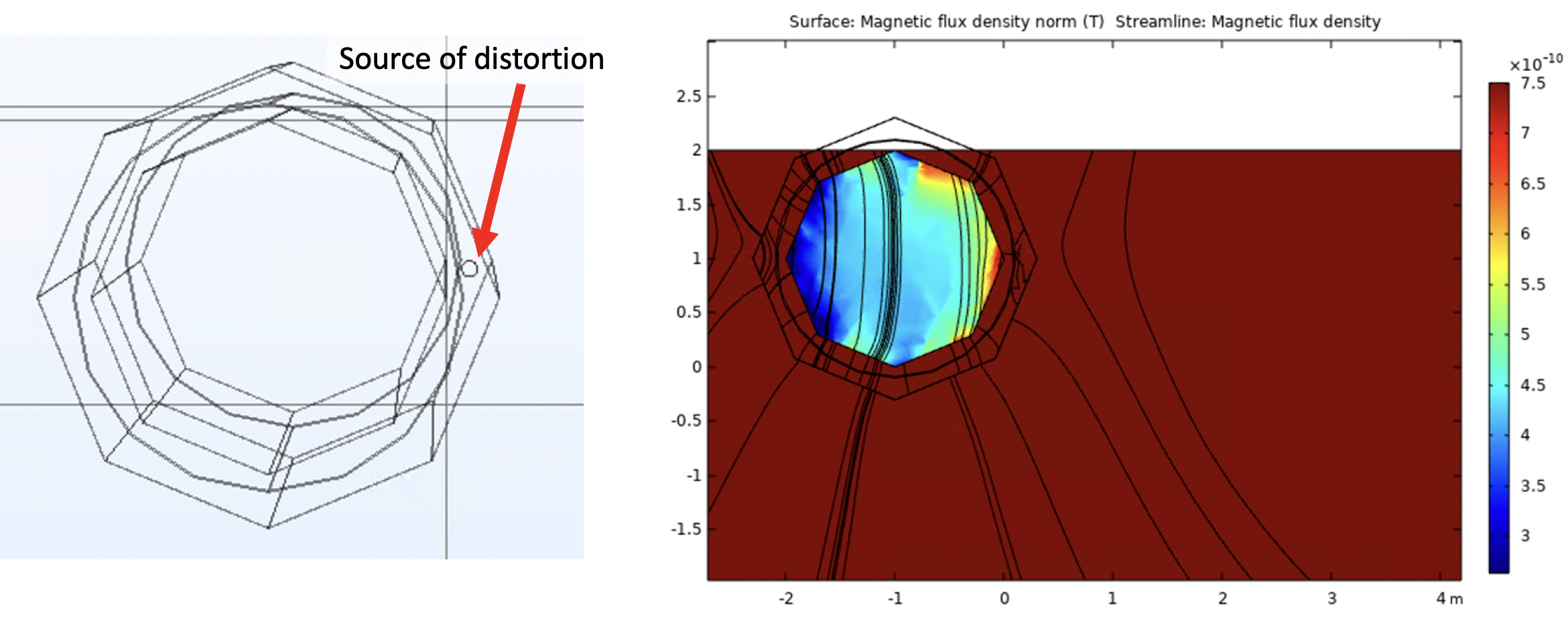}
    \caption{Example of a magnetic distortion due to a magnetic weld and impact on the inside field after degaussing. }
 \label{fig:distortions}
\end{figure}

Magnetic equilibration could be realized as in the simulation with 8 turns for each octagon, with 80~Amp-turns.
A further aspect are local magnetic distortions e.g. by the vacuum chamber or from the outside, which are modeled, see Figure~\ref{fig:distortions}. 
Here, a typical issue could be a magnetic weld in the vacuum chamber, which can be $>$~1000~nT over a distance of few centimeters.

A Monte Carlo study which tracked particles inside the shield using simulated residual field maps has then been performed. This showed that the efficiency loss due to the presence of a finite magnetic field is of the order of several per cent.

\subsection{Vacuum}
As explained previously the neutrons must be transported in a magnetically shielded and low vacuum region. For quasi-free neutrons, the vacuum pressure should be less than $10^{-5}$mbar \cite{vacuum1,vacuum2,vacuum3}. The vacuum requirements span a relatively big volume (approx. 2 m in diameter, 200 m long) connected directly to the ESS monolith vessel, imposing a very specific challenge from a mechanical, vibrational, vacuum, and radiation/safety engineering perspective. The vacuum vessel will contain the neutron optics for tens of meters see (Section \ref{sec:reflector}) (in vacuum) and will support the magnetic shield (external to the vacuum) for all the length of the beamline.\\
The ESS monolith was designed to be able to operate with He atmosphere (100K Pa) or at a low pressure $< \SI{1e-2}{Pa}$ range with a residual composition including He, \ce{H_2O}, and light hydrocarbons (CO, \ce{CO_2}, \ce{CH_4} and other fractions), as part of the expected environment of the spallation process with surface temperatures $> \SI{373}{K}$ and a mixed environment of neutrons and gamma radiation. 
A specific vacuum barrier will be necessary to offer the flexibility to work independently on the target and on the NNBAR experiment. A combination of mechanical pumps (dry rough and turbo-molecular) units for pump down and a combo-type (sputter ion pump and non-evaporable getter) are currently planned as a permanent pump solution  to minimise vibration on the optics system and assure a low level of physical access during the periods of operation. 

%% file: detector.tex
\section{The NNBAR detector}
\label{nnbardetector}


Design studies for the NNBAR detector using \software{GEANT4}\cite{Agostinelli:2002hh} are ongoing. These follow guiding principles based on the expected signal and background properties which are simulated with Monte Carlo-based models~\citep{Golubeva:2018mrz,Barrow:2019viz,Barrow:2021odz,CRY}. An overview of the design work is given in this Section, which is based on the detector model described in Ref.~\cite{sym14010076}. 

The goals of the work to date are the development of a detector model based on \software{GEANT4} together with sensitive variables for use in a search for neutron conversions. Future work includes the detailed quantification of background processes and sensitivity of the experiment. Here, the detector is described together with sensitive variables.

\subsection{Guiding Principles for Detector Design}
The conversion of a neutron to an antineutron together with the subsequent baryon number annihilation of an antineutron with a nucleon would be an extremely rare process. The NNBAR annihilation detector must therefore permit the observation of a tiny signal rate, to the level of a single signal event. There is therefore a need to suppress backgrounds to a negligible level. Reaching these goals drives the NNBAR detector design effort.

A common vertex of origin in the target foil from which several charged pion tracks arise can be reconstructed (to $\sim$mm precision). This requires excellent particle identification (PID). It must be possible to discriminate between charged pions from signal processes and protons from spallation-induced backgrounds. This can be achieved with measurements of the specific continuous energy loss, $\frac{dE}{dx}$. Time projection chambers (TPCs) are well suited to this task. Furthermore, it is important that the energies and directions of particles be measurable such that global annihilation event-level  quantities which are characteristic of the signal event, relating to signal properties such as isotropy, can be determined. 

Since the annihilation event happens in a nucleus, nuclear effects lead to the rejection power of some observables, such as the reconstructed final state invariant mass (which would naively be expected to be twice the neutron mass) being degraded. A consequence of this is that some of the requirements on the calorimeter energy reconstruction for the signal final-state particles can be relaxed. Since the antineutron-nucleon annihilation is a low energy process, a significant fraction of the available energy corresponds to the rest masses of the pions. For example, in a final-state comprising four pions, around 600~MeV of the total energy is accounted for in this way. High precision PID and  measurements of the multiplicities of different types of particles are therefore indirect energy measurements. In addition to tracking and the use of the $\frac{dE}{dx}$ technique,  precision electromagnetic calorimetry is required to find neutral pions via their decay $\pi^0\rightarrow \gamma \gamma$. 

Timing measurements, to order ns precision, from~a dedicated scintillator-based cosmic veto and from individual detector components, are required to suppress backgrounds from cosmic rays (charged and neutral). The~detector must also be sufficiently well granulated to be able to cope with the flux of gammas from neutron capture at the target and other beam-related backgrounds.

\subsection{Simulations}


Particle generation in signal processes is calculated with the model developed in Refs.~\citep{Golubeva:2018mrz,Barrow:2019viz,Barrow:2021odz}. The annihilation process at a particular radial position within the nucleus is simulated using a nuclear model of Fermi (gas) motion and taking into account an attractive antinucleon potential. Transport within the nucleus (final state interactions) is calculated with a stochastic intranuclear cascade model. Nuclear knockout processes leading to, e.g., protons emitted from the nucleus following nuclear transport and final state interactions, can also be modelled but correspond to particle production at lower kinetic energies than expected for signal pions. The simulation has been validated against low-energy interactions using antiproton data~\mbox{\citep{Barrow:2021odz,Golubeva:2018mrz}}.

The above model produces the expected characteristics from an annihilation event. An invariant mass of up to $\sim1.88\,$GeV (i.e. twice the nucleon mass) and final states of typically $3$-$6$ pions with momenta up to several hundred MeV are produced. The pions are roughly isotropically distributed for the annihilation channels used in this analysis. Kaon production during annihilation is heavily suppressed due to phase space effects and is not considered here. Resonances, such as $\eta$, $\rho$ and $\omega$ mesons, are included and are decayed by the model.

The CRY model~\cite{CRY} is used in this paper to study charged and neutral cosmic ray backgrounds. It produces a correlated cosmic-ray particle shower distributions at various elevations which is then used for the detector simulation studied in this paper. The CRY model is based on simulations of primary cosmic rays in the atmosphere and validated against cosmic-ray measurements. The simulation provides flux predictions for different types of particles (muons, neutrons, protons, electrons, photons, and pions). 

\subsection{The baseline NNBAR detector }\label{sec:baseline}
This detector model is implemented in \software{GEANT4} and includes a range of technology options. Crystal (CsI(Na)), lead-glass and sampling calorimeters (liquid argon/steel) have been studied. A TPC is similarly available, as well as a layer of silicon.       

The current baseline detector concept~\cite{sym14010076} comprises silicon tracking, placed within a 2cm thick aluminium beampipe. The TPC surrounds the beampipe. A hadronic range detector, comprising of 10 scintillator stacks, lies outside of the TPC. The slat orientation is arranged orthogonally for each layer, allowing rudimentary tracking with the scintillator stack.

These are surrounded by the electromagnetic calorimeter comprising lead-glass modules and a scintillator-based cosmic shield. The baseline concept allows a vertex reconstruction  to $\sim $mm precision using a silicon tracker and powerful particle identification via $\frac{dE}{dx}$ measurements in the TPC.

\begin{figure}[!htb]
     \centering
     \begin{subfigure}[b]{1\linewidth}
         \centering
         \includegraphics[width=\linewidth]{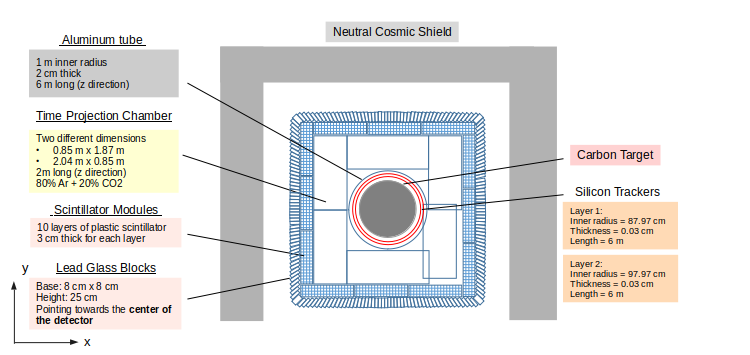}
         \caption{x-y view}
     \end{subfigure}
     \begin{subfigure}[b]{0.54\textwidth}
         \centering
         \includegraphics[width=\textwidth]{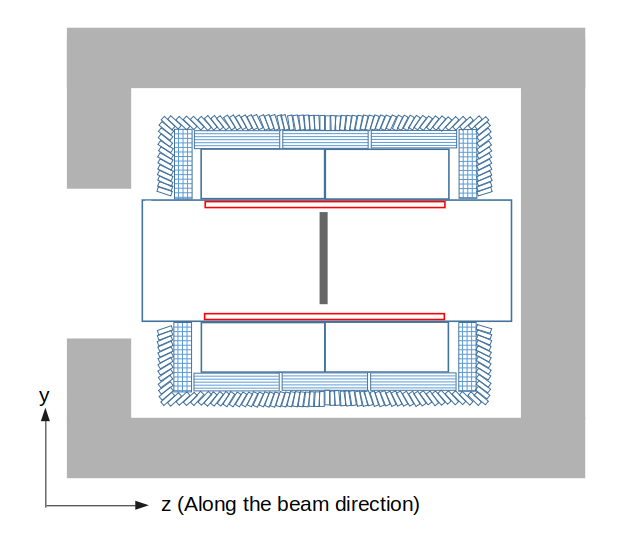}
         \caption{y-z view}
     \end{subfigure}
    \caption{Schematic overview of the NNBAR detector~design.}
    \label{box_full_det_sketch}
\end{figure}

Figure~\ref{box_full_det_sketch} shows an overview of the NNBAR detector. The detector is 600~cm long in the longitudinal ($z$) direction. In the transverse ($x-y$) plane, the detector has a width and height of 515~cm. The detector components and their dimensions are labelled in the figure. The annihilation target at the center is a  $100\,\mu$m thick carbon disk of $2\,$m in diameter.

In addition to the detector system, the detector is enclosed by a passive 1 meter thick lead shield to suppress cosmic ray backgrounds. Not shown in the Figure are arrays of scintillator tiles placed next to the passive shields.  



\subsubsection{Material budget}





\begin{figure*}[h]
\centering
\includegraphics*[width=0.55\textwidth]{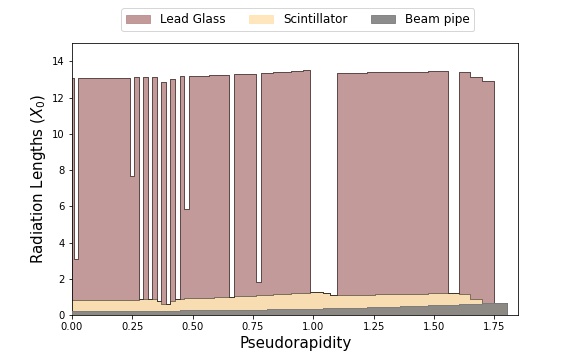} 
\caption{The radiation length, show cumulatively for each component, as a function of the absolute value of pseudorapidity.}
\label{fig:radiation_length_box}
\end{figure*}
\begin{figure*}[h]
\centering
\includegraphics*[width=0.55\textwidth]{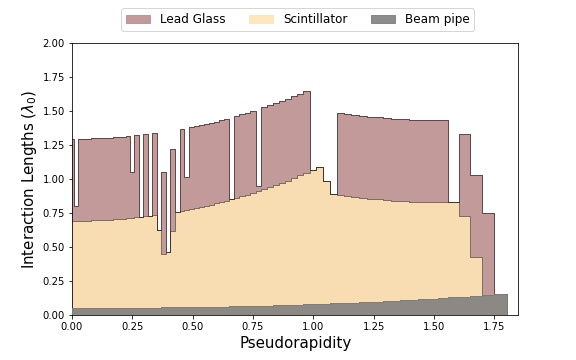} 
\caption{The interaction length, shown cumulatively for each component, as a function of the absolute value of pseudorapidity.}
\label{fig:interaction_length_box}
\end{figure*}

The computation of the material budget in the detector in terms of the radiation and interaction lengths is done in \software{GEANT4}. The contributions of radiation length ($X_0$) and interaction length ($\lambda_0$) by each detector component as a function of the absolute value of pseudo-rapidity\footnote{The pseudorapidity ($\eta=-ln(tan{\frac{\theta}{2}})$) is defined with respect to the polar angle $\theta$ where $\theta=0$ denotes the $+z$-direction i.e. along the centre of the beam pipe towards the detector through the magnetically shielded region.}~are shown in Figure~\ref{fig:radiation_length_box} and~\ref{fig:interaction_length_box} respectively and extend up to but do not include the active and passive cosmic ray shields.  The lead glass blocks correspond to around 11$X_0$ to ensure good energy containment. In total, the detector corresponds to $1-1.5\lambda_0$.  

Since the silicon detectors are thin, they have negligible contribution to both the radiation length and interaction length. It can also be seen from Figure~\ref{fig:radiation_length_box} and~\ref{fig:interaction_length_box} that there are regions where there is no radiation length contribution from the lead glass blocks. This is due to the gaps between lead glass blocks and an underlying assumption for the purpose of producing Figure~\ref{fig:radiation_length_box} and \ref{fig:interaction_length_box} that particles originate from the center of the target foil. A larger gap at pseudo-rapidity $\sim 1$ indicates the location of the detector edge. 



\begin{figure}[!htb]
     \centering
     \begin{subfigure}[b]{0.48\textwidth}
         \centering
         \includegraphics[width=\textwidth]{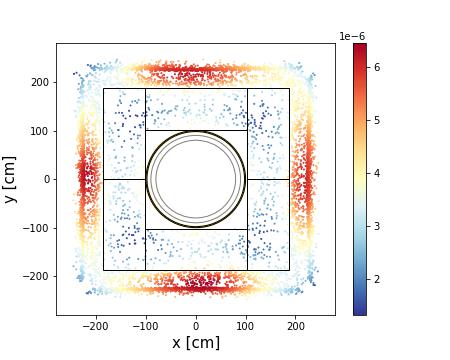}
         \caption{x-y view}
     \end{subfigure}
     \begin{subfigure}[b]{0.48\textwidth}
         \centering
         \includegraphics[width=\textwidth]{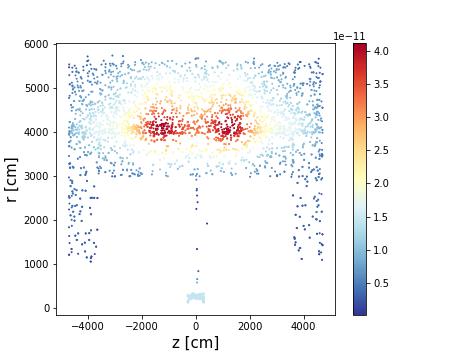}
         \caption{r-z view}
     \end{subfigure}
    \caption{Positions of $\gamma \rightarrow e^+ + e^-$ conversions in and around the detector.}
    \label{fig:gamma conv map}
\end{figure}

A photon from a neutral pion decay can undergo conversion to an electron-positron pair before reaching the lead glass. If a photon converts before reaching the calorimeter, a single highly ionising charged track could be left in the TPC since there is no magnetic field in the detector and $e^+e^-$ pairs are produced with a tiny opening angle. The probability of conversion processes in different detector components can be estimated by simulating $\pi^0$ decays from the signal events. Figure \ref{fig:gamma conv map} shows the locations where photons convert the in $x-y$ and $r-z$ views. Results from the \software{GEANT4} simulation shows that $\sim 18\%$ and $\sim 29\%$ of the photons convert in the beam pipe and the  scintillator stacks respectively. Around $1\%$ of the photons convert at silicon trackers. The remainder ($\sim 50\%$) can be measured in the lead-glass calorimeter if they are in the appropriate acceptance region. 

\subsection{Trigger} 
The main goal of the NNBAR detector trigger system is the ability to identify and read out neutron-antineutron annihilation candidate events. To~minimize overlaid detector hits from unrelated background events the trigger must be able to resolve accepted events in a short time window. To this end, the NNBAR experiment will employ a two-level trigger system. The first level will be hardware based and work in self-trigger mode of individual calorimeter cells and cosmic veto sensors. 
The required readout bandwidth for the subdetectors is expected to be relatively modest since only self-triggered calorimeter and veto channels will be read-out.
This may allow the use of a commercial network switch for connecting the front-end readout with the off-detector processor farm, with PMT samples and timestamp data encoded in compatible network packet formats.
Each channel will be comprised of a global time stamp which will allow the timing information of different subdetectors to be matched consistently. The time stamps are to be reconstructed to ns-level precision. Moreover, high-speed sampling analog-to-digital converters (ADCs) will produce digitized samples of the shaped analog pulses allowing the pulse amplitudes of individual channels to be reconstructed.
The second level of the trigger will be implemented off-detector in a dedicated CPU farm. The time-stamped readout of the hardware trigger will be used to select event candidates and the corresponding TPC data frames for further event reconstruction and storage. 
It is worth mentioning that in the ILL experiment \cite{Baldo-Ceolin:1994hzw}, the long trigger time window (150 ns) caused a pile-up of multiple gamma events from neutron capture in the carbon target to be accepted (individual gammas can reach energies as high as $\sim$5 MeV). This accounted for as much as 32 \% of the total trigger rate ($\sim$4 Hz). With ~fast scintillator and/or Cherenkov detectors read out with high-speed ADCs the~time windows will be significantly shortened, which should substantially reduce the acceptance of multiple piled up gammas.

The trigger time window of the NNBAR detector system, even in the worse case scenario, will be reduced by a factor of three compared to the ILL experiment, if the trigger timing resolution would be limited to the ADC clock interval. In such case the the~maximum trigger window would be double the ADC clock period (e.g., 50 ns for a 40 MHz ADC rate similar to the Large Hadron Collider~\cite{Evans:2008zzb}) due to the arbitrary timing of the arriving neutrons. Modern trigger/DAQ system will allow fine timing extraction from the shaped, digitized pulses, which should further reduce the trigger timing windows.

The trigger acceptance criteria in the simplest scenario will be based on events with multiple hits in the scintillator layers above a minimum threshold. Additional trigger criteria based on particle identification and event topology may further improve selectivity.

\subsection{Cosmic ray shielding} 
\begin{figure}
     \centering
     \begin{subfigure}[b]{0.48\textwidth}
         \centering
         \includegraphics[width=\textwidth]{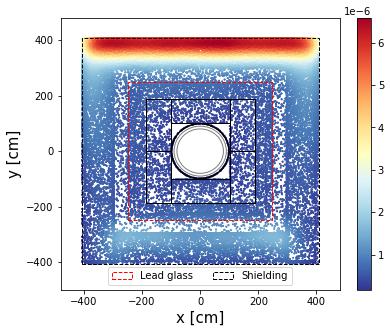}
         \caption{}
     \end{subfigure}
     \begin{subfigure}[b]{0.48\textwidth}
         \centering
         \includegraphics[width=\textwidth]{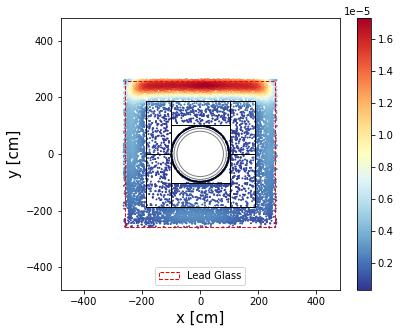}
         \caption{}
     \end{subfigure}
    \caption{Cosmic stopping position (a) with and (b) without the shielding system.}
    \label{fig:cosmic stop pos}
\end{figure}

The shielding effect of cosmics ray background is also studied. The \software{CRY} generated cosmic ray background particles are tracked and their stopping positions are recorded. The stopping position is defined by the space point where the primary particle comes to rest in the simulation. Hence, only particles fully stopped in the detector and shielding are taken into account. The result is shown in Figure \ref{fig:cosmic stop pos}. In the case of detector with overburden shielding, the majority of the neutral cosmics ray background will be stopped in the passive shielding, while most of the muons will punch through. The portion of neutral cosmic background particles stopping in the shield, in different energy ranges, are listed in Table \ref{table:cosmic_stopping_percentage}.

\begin{table}[h!]
    \begin{centering}
        \begin{tabular}{|c|c|c|} 
        \hline
        KE(GeV) & Neutron fraction & Photon fraction \\
        \hline
        0-0.5 & 0.97 & 0.97 \\
        0.5-1 & 0.97 & 0.97\\
        1-5   & 0.96 & 0.97\\
        5-10  & 0.83 & 0.84 \\
        10-50 & 0.87 & 0.87 \\
        $>$50 & 0.88 & 1.00 \\
        \hline
        \end{tabular}
        \caption{The fraction of neutral cosmic ray particles stopped in the detector for different ranges of kinetic energy (KE), as simulated using \software{CRY}. }
        \label{table:cosmic_stopping_percentage}
    \end{centering}
\end{table}

\subsection{Sensitive variables}\label{sec:evsel}
Particle and event variables can be used for characterizing antineutron-nucleon annihilation events and discriminating against backgrounds. Such quantities include 
the invariant mass ($W$) and sphericity ($S$) of the observed particle system in the detector. Their computation uses algorithms to identify and reconstruct particles (neutral and charged pions and protons)~\cite{sym14010076}. An example of another sensitive variable is the difference in timing between the first and last scintillator hits ($\Delta T$).

\begin{figure}
     \centering
     \begin{subfigure}[b]{0.45\textwidth}
         \centering
         \includegraphics[width=\textwidth]{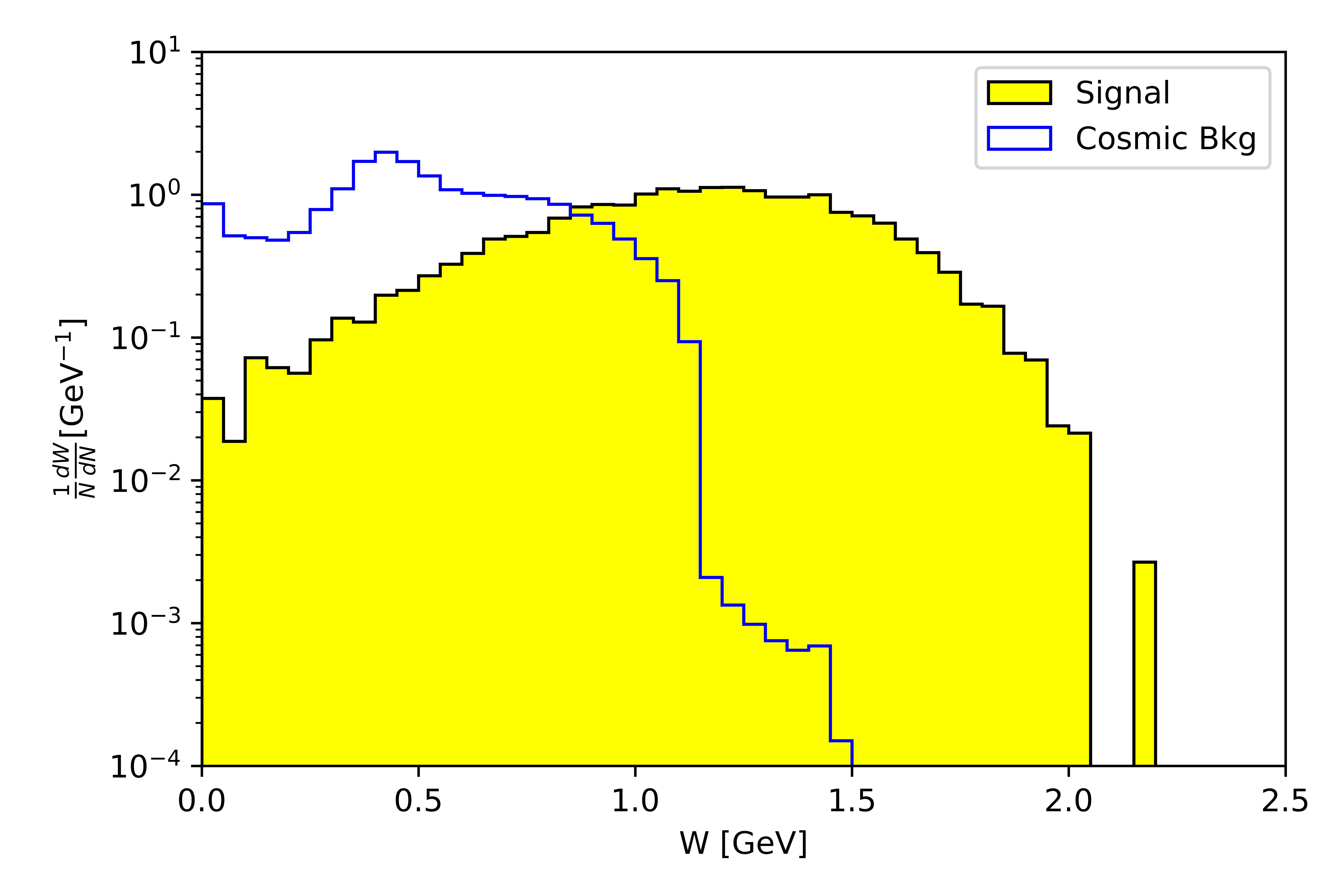}
         \caption{Invariant Mass $W$.}
     \end{subfigure}
     \begin{subfigure}[b]{0.45\textwidth}
         \centering
         \includegraphics[width=\textwidth]{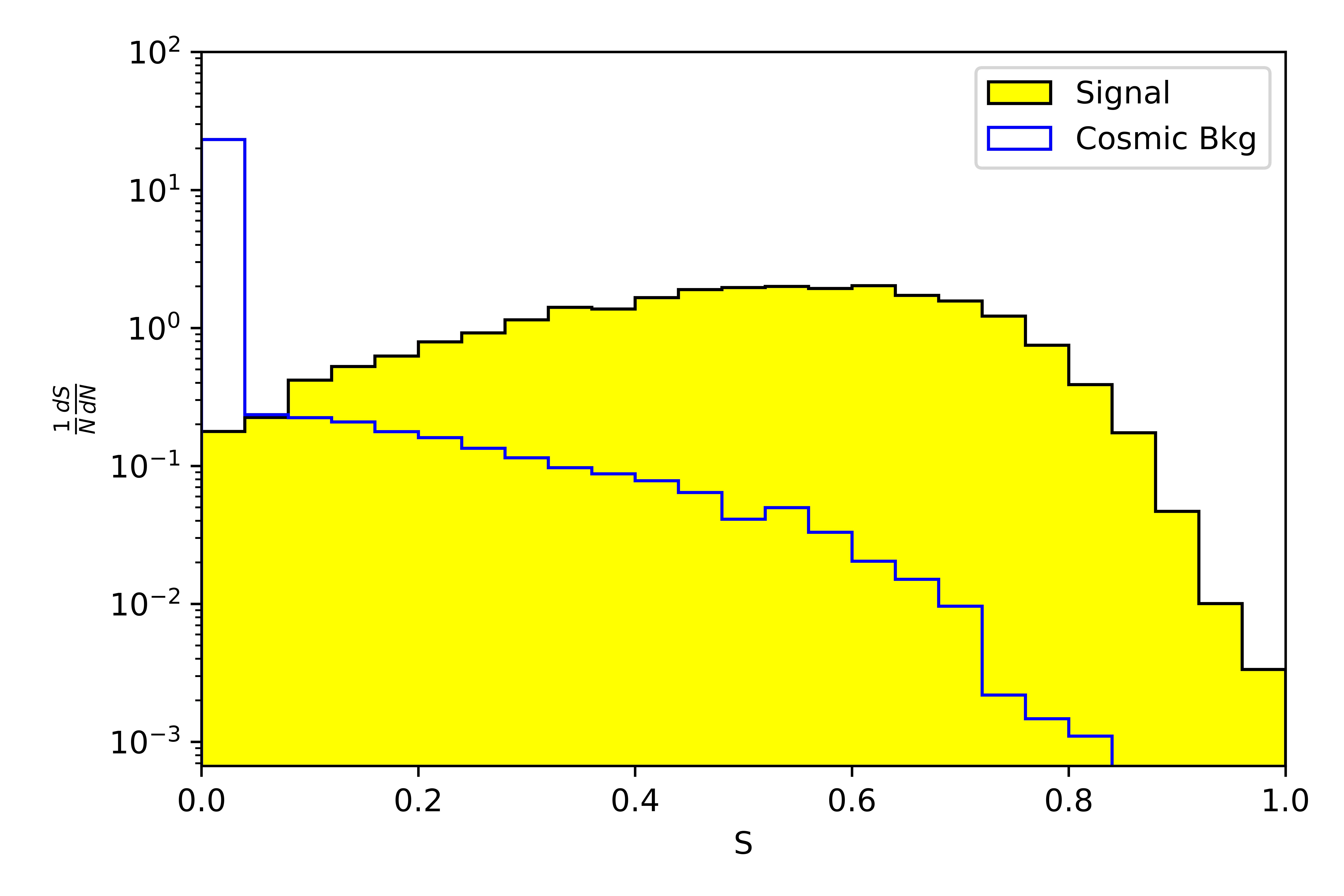}
         \caption{Sphericity S}
     \end{subfigure}
     \begin{subfigure}[b]{0.45\textwidth}
         \centering
         \includegraphics[width=\textwidth]{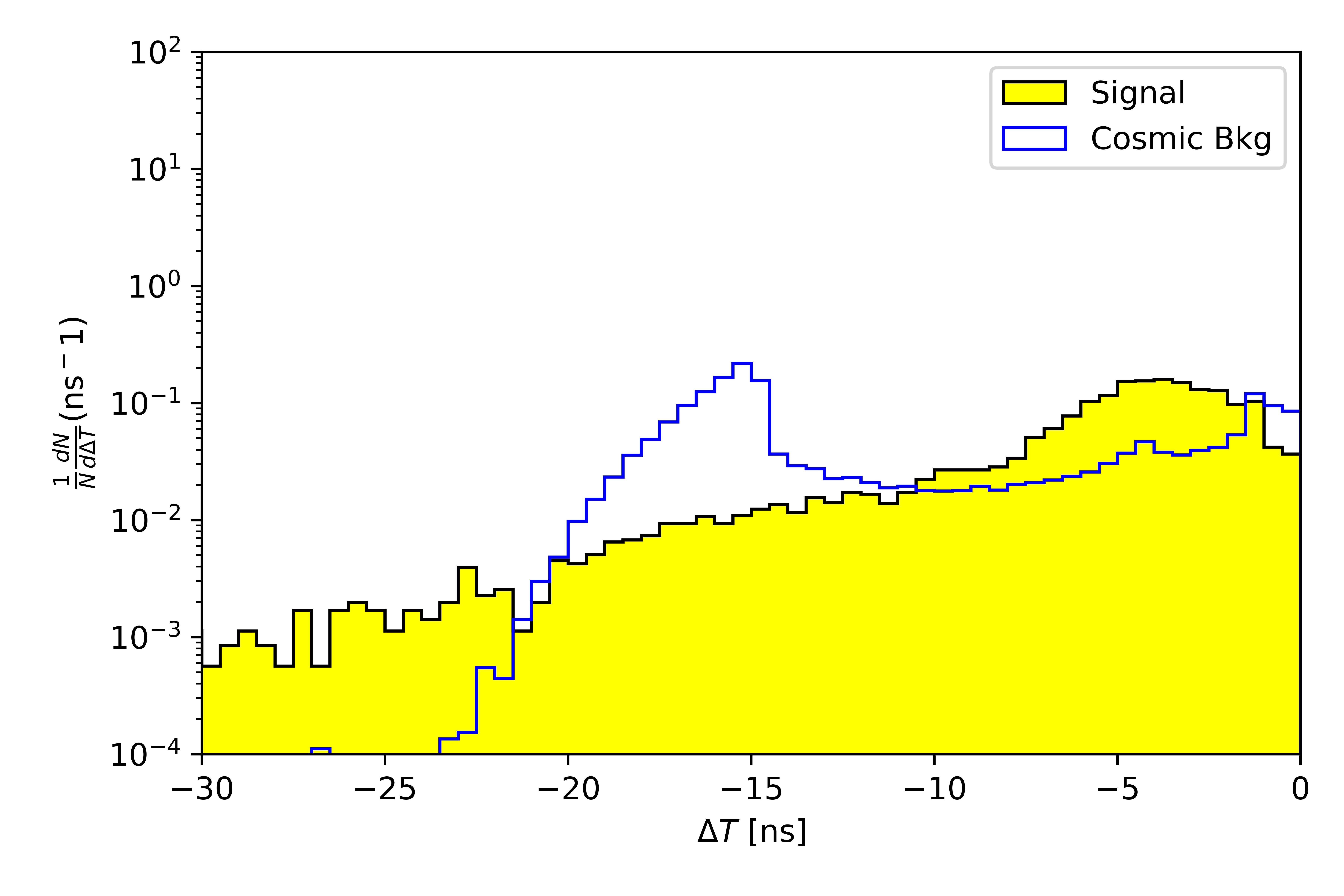}
         \caption{Timing $\Delta T$.}
         \label{subfig:timing_var}
     \end{subfigure}
    \caption{Distributions of $W$, $S$ and $\Delta T$ for expected signal and cosmic ray backgrounds.}
    \label{fig:event_var_dist}
\end{figure}

Figure \ref{fig:event_var_dist} shows event variable distributions of the signal events and cosmic ray background events. The cosmic ray background events are composed of simulation samples from cosmic muons, neutrons, and photons. Clear discrimination between signal and cosmic ray backgrounds can be seen for each distribution. A basic event selection algorithm has been developed to suppress cosmic ray backgrounds.



\subsection{Alternative detector choices}

\begin{figure}[tb]
	\centering
	\includegraphics*[width=0.725\textwidth]{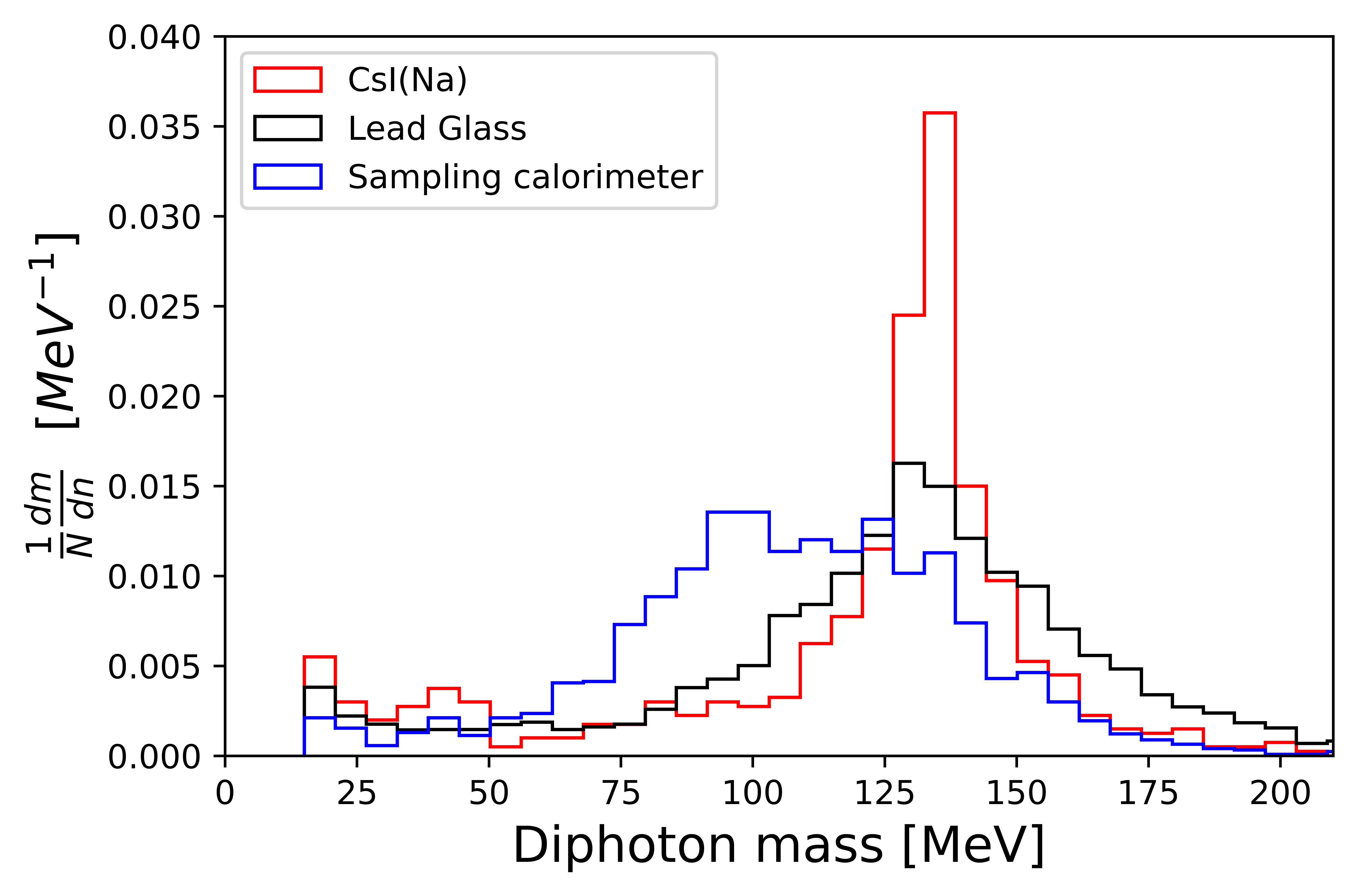}
	\captionsetup{width=0.99\linewidth}
	\caption{Diphoton mass spectrum for single pion events for a crystal, lead-glass and sampling calorimeter.}
	\label{fig:pi0}
\end{figure}

Although the results of the baseline detector have been shown here, the model has the flexibility to incorporate different approaches and technology choices. An example of this is given here with respect to neutral pion reconstruction - the observation of a neutral pion is expected to be one of the key discriminants between signal and background events.  Figure~\ref{fig:pi0} shows the expected diphoton spectrum for events containing a neutral pions with 100 MeV kinetic energy for a crystal (CsI(Na)), lead-glass and sampling calorimeter (42 layers of liquid argon and stainless steel). Each calorimeter has the same cell segmentation and size as implemented for the baseline (lead-glass) system. No correction is made for energy not deposited in the liquid argon for the sampling calorimeter. 

As would be expected, the crystal calorimeter delivers the superior resolution, followed by lead-glass and the  sampling calorimeter. However, lead-glass is so far chosen as the baseline option owing to cost (crystal calorimetry may be prohibitively expensive for a large detector), appropriate resolution for $\pi^0$ identification and background rejection (heavy nuclear fragments would not pass the Cherenkov threshold and thus would not be recorded as for other calorimeter choices). 

 The subsequent stages of the work involve the optimisation of the detector model using a range of predicted backgrounds in order to develop a cost-effective detector solution.

%% file: summary.tex
\section{Summary}
\label{sec:summary}
The NNBAR instrument for the European Spallation Source will look for neutrons transforming to antineutrons with a sensitivity improvement of three orders of magnitude compared with earlier searches. This paper describes the development of the design of the experiment. It focuses on each of the critical components needed - a cold moderator, high precision neutron focusing, field-free propagation along a dedicated beamline and detection of the annihilation signal. Each of these studies form parts of a model of the experiment which will be used to make a final quantification of the experiment's cost and physics potential in a conceptual design report. 

\section{Acknowledgements}
This work was funded by the HighNESS project at the European Spallation Source. HighNESS is funded by the European Framework for Research and Innovation Horizon 2020, under grant agreement 951782. Support was also given by the Swedish Research Council, Vetenskapsr{\aa}det. 

%% file: acknowledgements.tex
\label{Acknowledgements}
{

%% file: nnbar.bbl
\providecommand{\href}[2]{#2}\begingroup\raggedright\endgroup